\documentclass{aa}

\usepackage{graphicx}
\usepackage{txfonts}
\usepackage{lipsum}
\usepackage{lscape}
\usepackage{placeins}
\usepackage{amsmath}
\usepackage{amssymb}
\usepackage{bm}
\usepackage{multirow}
\usepackage{longtable}
\usepackage{booktabs}
\usepackage{makecell}
\usepackage{color}
\usepackage{xcolor}
\usepackage{url}
\usepackage{appendix}
\usepackage[colorlinks=true, linkcolor=blue, citecolor=blue, urlcolor=blue]{hyperref}

\newcommand{\re}[1]{\textcolor{black}{#1}}
\newcommand{\ree}[1]{\textcolor{black}{#1}}

\graphicspath{{./}{figures/}}
\makeatletter
\@ifundefined{linenumbers}{}{%
  \let\linenumbers\relax

  \let\nolinenumbers\relax
}
\makeatother

\begin{document}
   \title{Planetary edge trends}
   \subtitle{I. The inner edge \text{--} stellar mass correlation}
   \author{Meng-Fei Sun\inst{1,2}
        \and Ji-Wei Xie\inst{1,2}
        \and Ji-Lin Zhou\inst{1,2}
        \and Beibei Liu\inst{3}
        \and Nikolaos Nikolaou\inst{4}
        \and Sarah C. Millholland\inst{5,6}
        }
   \institute{School of Astronomy and Space Science, Nanjing University, Nanjing 210023, PR China\\
             \email{jwxie@nju.edu.cn}
            \and Key Laboratory of Modern Astronomy and Astrophysics, Ministry of Education, Nanjing 210023, PR China
            \and Institute for Astronomy, School of Physics, Zhejiang University, Hangzhou 310027, PR China
            \and Department of Physics and Astronomy, University College London, Gower Street, WC1E 6BT London, United Kingdom
            \and Department of Physics, Massachusetts Institute of Technology, Cambridge, MA 02139, USA
            \and Kavli Institute for Astrophysics and Space Research, Massachusetts Institute of Technology, Cambridge, MA 02139, USA\\}
   \date{Received September 30, 20XX}
 
  \abstract
   {Recent advancements in exoplanet detection have led to over 5,900 confirmed detections. The planetary systems hosting these exoplanets exhibit remarkable diversity.}
   {The position of the innermost planet (i.e., the inner edge) in a planetary system provides important information about the relationship of the entire system to its host star properties, offering potentially valuable insights into planetary formation and evolution processes.} 
   {In this work, based on the $Kepler$ Data Release 25 catalog combined with LAMOST and $Gaia$ data, we investigate the correlation between stellar mass and the inner edge position across different populations of small planets in multi-planetary systems, such as super-Earths and sub-Neptunes. By correcting for the influence of stellar metallicity and analyzing the impact of observational selection effects, we confirm the trend that as stellar mass increases, the position of the inner edge shifts outward.}
   {Our results reveal a stronger correlation between the inner edge and stellar mass ($a_{\text{in}} \propto M_{\star}^{\gamma_1}$), with a power-law index of $\gamma_1 = 0.6-1.1$, which is larger compared to previous studies. The stronger correlation in our findings is primarily attributed to two factors: first, the metallicity correction applied in this work enhances the correlation; second, the previous use of occurrence rates to trace the inner edge weakens the observed correlation.}
   {Through comparison between observed statistical results and current theoretical models, we find that the pre-main-sequence dust sublimation radius of the protoplanetary disk best matches the observed inner edge\text{--}stellar mass. Therefore, we conclude that the inner dust disk likely limits the innermost orbits of small planets, contrasting with the inner edges of hot Jupiters, which are associated with the magnetospheres of gas disks, as suggested by previous studies. This highlights that the inner edges of different planetary populations are likely regulated by distinct mechanisms.}


   \maketitle

\section{Introduction} 
\label{intro}

The emergence of exoplanet surveys, such as the $Kepler$ mission, has revolutionized our understanding of the diversity and abundance of planetary systems \citep{2016RPPh...79c6901B}. With the advancement of exoplanet detection, over 5,900 exoplanets have been confirmed to date,\footnote{\url{https://exoplanetarchive.ipac.caltech.edu/index.html}} and these exoplanets reside in planetary systems that exhibit a remarkable diversity \citep{2021ARA&A..59..291Z}. The uniqueness of our Solar System has long been a topic of interest \citep{2018ApJ...860..101Z}. For instance, Mercury, the innermost planet in the Solar System, orbits the Sun at a distance of about 0.4 AU. In contrast, many planetary systems host planets that orbit much closer to their host stars, within 0.4 AU \citep{2018AJ....156...24M}. \re{Given the diverse planetary systems observed, an important question arises regarding how our own Solar System should be explored, understood, and interpreted.} To address this, we aim to investigate the intrinsic patterns reflected by the inner edges of planetary systems. Specifically, we focus on the positions of the innermost planets. Gaining a deep understanding of the intricacies of where these planets reside within their respective systems \citep{2022AJ....164...72M,2023ApJ...954..137S} will provide invaluable insights into the fundamental processes that govern planetary formation, migration, and stability. In recent years, a wealth of observational and theoretical research has illuminated the complex relations between stellar properties and planetary architectures \citep{2015ARA&A..53..409W,2021ARA&A..59..291Z,2023ASPC..534..863W}. Therefore, studying the relationship between the distribution of the inner edge and stellar properties is crucial.

The inner edge of protoplanetary disks around pre-main-sequence solar-type stars is located at approximately 0.1 AU \citep{2007prpl.conf..539M}, which corresponds with findings in several statistical studies that observed clustering of the inner edges of planetary systems around a 10-day (0.1 AU) orbital period \citep{2017ApJ...842...40L,2018AJ....156...24M}. Some theoretical studies have analyzed the mechanisms that determine the inner edge. For in situ formation, the location of the inner edge of the protoplanetary disk is related to the dust sublimation radius \citep{2001ApJ...560..957D,2008ApJ...673L..63P,2009A&A...506.1199K,2011Icar..212..416M,2017ApJ...845...44T,2019A&A...632A...7L}. Due to the lack of solid material required for in situ formation \citep{2013arXiv1306.0566B}, planets are less likely to be formed within these radii \citep{2013MNRAS.431.3444C}. For disk migration, planets form at more distant locations and migrate inward through the gas disk, becoming trapped at the co-rotation radius or the sublimation radius \citep{1996Natur.380..606L,2006ApJ...652..730M,2012ARA&A..50..211K,2013ApJ...764..105S,2023ApJ...951L..19B}. As a result, it is less likely for planets to exist within these radii \citep{2002ApJ...574L..87K}. Stellar tides and planetary tides within multi-planetary systems can also affect the radii where there is a lack of planets \citep{2007ApJ...670..820W,2009ApJ...698.1357J}. However, it remains unclear which theoretical model plays a dominant role, and further observational constraints and testing are needed to determine the most plausible model. \re{In this work, we summarize the theoretical models discussed above.}

As observational data continue to increase, using large datasets such as those from $Kepler$ to test inner-edge theoretical models has become possible. Several observational studies have investigated the relationship between the inner edge of planetary systems and stellar mass using a power-law model: $a_{\text{in}} \propto M_{\star}^{\gamma_1}$, where the power-law index $\gamma_1$ represents this correlation (as in our work). \cite{2013ApJ...769...86P} analyzed exoplanets orbiting within 0.1 AU, including confirmed Jovian exoplanets and $Kepler$ planetary candidates from the early data release (i.e., the third tabulation of $Kepler$ planetary candidates). Their findings confirmed a correlation between the planetary migration halting distance (i.e., the inner edge) and stellar mass, which they examined through this power-law model. They found that migration halting from tidal circularization mechanism \citep{2006ApJ...638L..45F,2007ApJ...670..820W} provided the best fit for the empirical distribution of confirmed Jovian exoplanets. However, for $Kepler$ candidates, they derived a power-law index of $\gamma_1 = 0.38-0.9$, which exceeds the tidal halting mechanism’s prediction of $\gamma_1 = 0.23-0.33$. \cite{2015ApJ...798..112M} analyzed the entire $Kepler$ sample of smaller planets to assess planet occurrence around stars of varying spectral types, revealing notable differences across these types. The occurrence drop depended on stellar mass and scaled with the semimajor axis with a power-law index of $\gamma_1 \sim 0.33$. Their findings indicated that this scaling closely aligned with the pre-main-sequence co-rotation radius, leading them to derive a smaller power-law index for the planet occurrence\text{--}stellar mass scaling law. \cite{2024A&A...686L...1M}, by combining TESS and $Gaia$ DR3 data, identified a sample of 47 intermediate-mass main-sequence stars hosting confirmed and strong candidate hot Jupiters. Their findings indicated that hot Jupiters around intermediate-mass stars tended to orbit closer to the central stars than the inner dust disk, generally aligning with the magnetospheric truncation radius. Their result suggested that the inner gas disk, rather than the dust disk, constrained the innermost orbits of hot Jupiters around intermediate-mass stars. 

As shown by observations, the relationship between the inner edge and stellar properties, especially stellar mass, remains inadequately understood and warrants further investigation. This complexity may stem from differences in planetary populations, including variations in size. Planetary systems hosting giant planets versus small planets may exhibit different relationships between the inner edge and stellar mass due to differing formation and evolution processes \citep{2020RAA....20..164L,2023ASPC..534..717D}. Additionally, small planets themselves are highly diverse, with significant differences between super-Earths and sub-Neptunes \citep[e.g.,][]{2022Sci...377.1211L,2024A&A...687A..25C}. To clarify the relationship between the inner edge and stellar mass, a detailed analysis of planets of different sizes is required. Furthermore, stellar mass is also correlated with other properties, such as stellar metallicity, which itself plays an important role in planet formation and evolution \citep[e.g.,][]{2021ARA&A..59..291Z} and thus influences the inner edge. \re{Previous statistical studies have demonstrated the significant impact of metallicity on planetary system architectures \citep{2013ApJ...763...12B,2016AJ....152..187M}. Therefore, to} determine the dependence of the inner edge on stellar mass, it is essential to remove the effects of stellar metallicity. 

In this paper, to address these issues, we systematically analyze the relationship between the inner edge of planetary systems and the stellar mass by leveraging comprehensive and enriched observational data from LAMOST \citep{2021ApJ...909..115C,2021AJ....162..100C}, $Gaia$ \citep{2020AJ....159..280B}, and $Kepler$ \citep{2018ApJS..235...38T}. By combining LAMOST’s spectroscopic measurements with $Gaia$’s astrometric data, we are able to achieve precise characterization of $Kepler$ host stars, offering a large and homogeneous sample with accurate stellar properties. For example, the uncertainty in stellar mass is only 7\% \citep{2020AJ....159..280B}. \re{Although this is the best available stellar mass data for this study, it is important to note that the 7\% refers to precision rather than accuracy.} Furthermore, LAMOST provides a large sample of host stars with metallicity measurements \citep{2014ApJ...789L...3D}, which we utilize to investigate the impact of stellar metallicity on the inner edge\text{--}stellar mass relationship. Given that metallicity may influence the inner edge \citep{2016AJ....152..187M} and is highly correlated with stellar mass \citep{2010PASP..122..905J}, understanding its impact on the dependence of the inner edge on stellar mass is essential. Additionally, motivated by the radius gap observed in $Kepler$ systems \citep{2013ApJ...775..105O,2017AJ....154..109F}, we further divide planetary systems into super-Earth, sub-Neptune, and mixed systems for a more detailed analysis and to better reflect the relationship between the inner edge and stellar mass. 

This work is organized as follows: In Sect. \ref{sec:dataandselec}, we introduce the samples used in this study and the process of selecting both stellar and planetary samples. Then, in Sect. \ref{sec:anaandres}, we present the results of our statistical analyses to explore the observational findings. In Sect. \ref{sec:impanddis}, we compare and discuss our findings with theoretical models, extend our analysis by comparing our results with previous studies, and provide insights for future work. Finally, in Sect. \ref{sec:sumandcon}, we summarize the results and conclusions of this study and provide perspectives for future research. Additional detailed discussions and analyses are included in the appendix \ref{Mulders2015}.

\section{Data and sample selection}
\label{sec:dataandselec}

\subsection{Dataset}
\label{sec:datasamples}

The planetary properties are sourced from the $Kepler$ Data Release 25 (DR25) catalog \citep{2018ApJS..235...38T}, which includes 8,054 $Kepler$ Objects of Interest (KOIs). The primary stellar properties we analyzed, such as stellar mass, are based on the \cite{2020AJ....159..280B} $Gaia$-$Kepler$ Stellar Properties Catalog (hereafter referred to as Berger20), which includes 186,301 stars. Stellar metallicity data is derived from PAST \uppercase\expandafter{\romannumeral2} \citep{2021AJ....162..100C}, utilizing the LAMOST-$Gaia$-$Kepler$ Catalog (hereafter referred to as PAST \uppercase\expandafter{\romannumeral2}), which contains 35,835 stars and 1,060 planets. Systems lacking metallicity data (1,294 stars) were removed from our analysis. By combining data from $Kepler$ DR25, Berger20, and PAST \uppercase\expandafter{\romannumeral2}, we established the dataset for this study as shown in Table \ref{tab:dataselection}.

\begin{table}[htbp]
\centering
\renewcommand\arraystretch{1.5}
\caption{\centering Data selection.}
\label{tab:dataselection}
{\footnotesize
\linespread{1.9}
\begin{tabular}{l|cc}
\hline\hline
Match and Filter & Star & Planet \\\hline
$Kepler$ DR25 & 6923 & 8054 \\
Match with $Gaia$ (Berger20) & 6391 & 7467 \\
Match with LAMOST (PAST \uppercase\expandafter{\romannumeral2}) & 1420 & 1749 \\
Not a false positive & 727 & 1025 \\
Not a binary star & 727 & 1025 \\
Not a giant star & 708 & 1003 \\
P $>$ 1 day & 688 & 980 \\
R$_{\rm p}$ $<$ 4 R$_\oplus$ & 591 & 835 \\
S/N $>$ 7.1 & 587 & 829 \\
N$_{\rm p}$ $>$ 1 & 166 & 408 \\\hline
\end{tabular}}
\tablefoot{N$_{\rm p}$ is the number of transiting planets in a system during the process of data and sample selection in Sect. \ref{sec:dataandselec}.}
\end{table}

\begin{figure*}[htbp]
\setlength{\abovecaptionskip}{-0.3em}
\centering
\begin{minipage}[b]{0.45\textwidth}
    \centering
    \includegraphics[width=\textwidth]{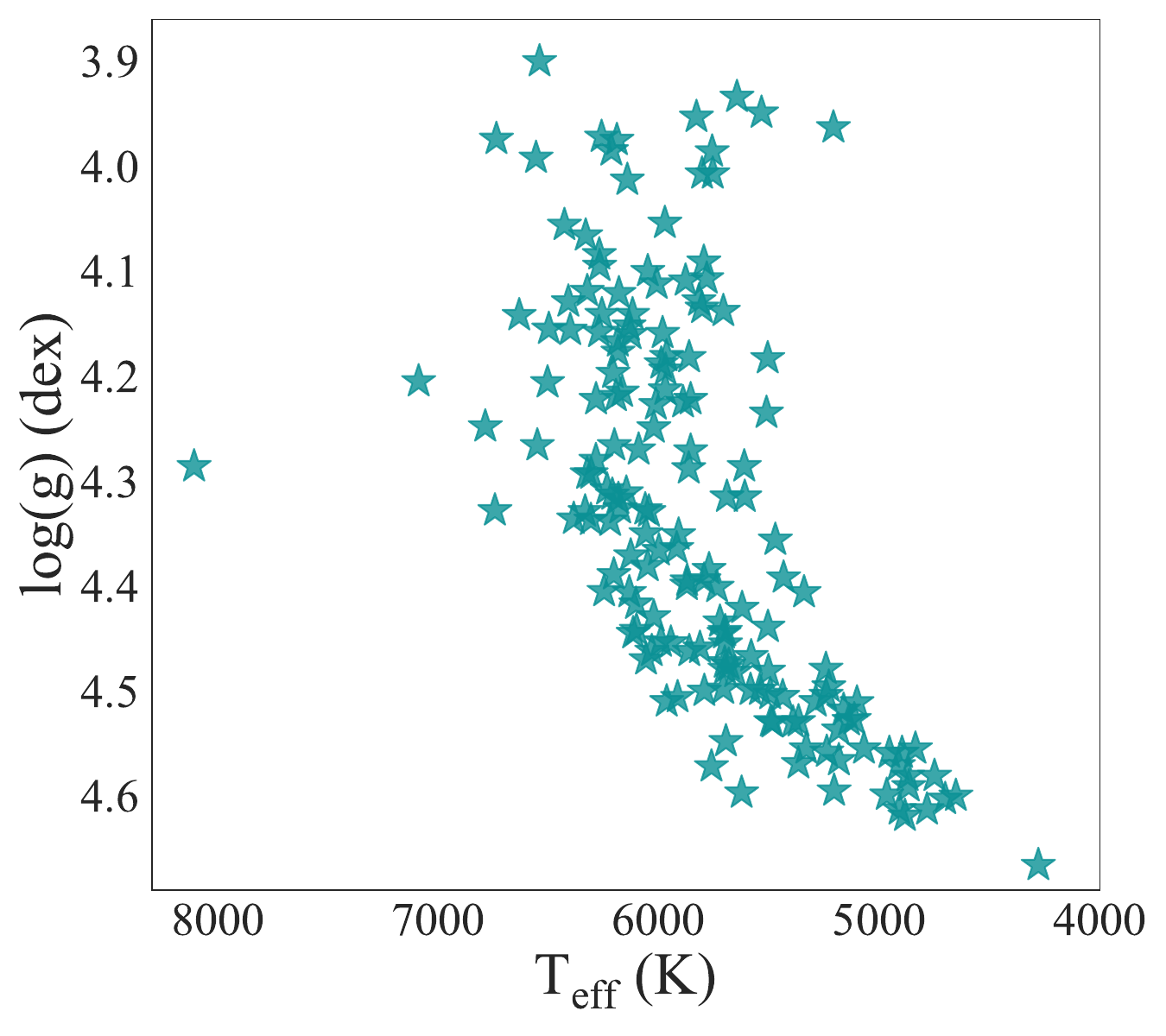}
\end{minipage}
\begin{minipage}[b]{0.45\textwidth}
    \centering
    \includegraphics[width=\textwidth]{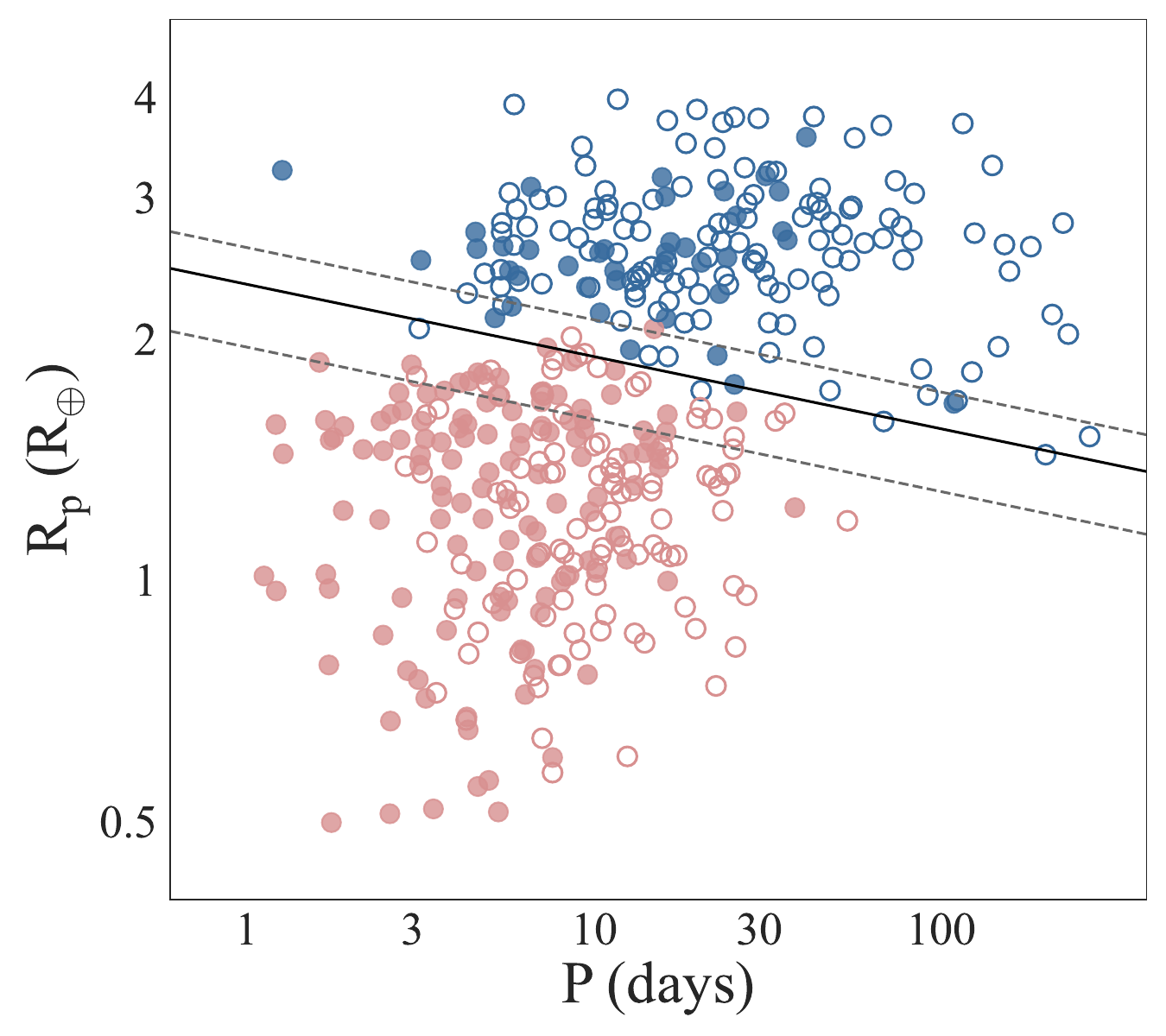}
\end{minipage}
\vspace{-0.6em}
\caption{Distribution of stars and planets for all multis studied in this paper. The left panel illustrates the distribution of stars, while the right panel presents the distribution of planets. Solid points denote the inner edge of the system, while hollow points represent other planets within the system. Red points indicate super-Earths and blue points represent sub-Neptunes, classified according to the radius valley defined in Eq. (\ref{radiivalley}). The black solid line indicates $M_\star = 1.0 \, M_\odot$, and the upper and lower dashed lines correspond to $M_\star = 1.5 \, M_\odot$ and $M_\star = 0.5 \, M_\odot$, respectively.}
\label{figdata2HRandPR}
\end{figure*}

\subsection{Stars and planets}
\label{sec:sampleselection}

We used the data filtering process from Table \ref{tab:dataselection} as the standard case and selected stars and planets according to the following filters: 
\begin{enumerate}
    \item Not a false positive. We excluded planets flagged as false positives from our analysis.
    \item Not a binary star. The renormalized unit weighted error (RUWE) provides a renormalized metric that is independent of color and magnitude, and it helps identify and exclude binary systems. Therefore, we used RUWE $<$ 1.2 to exclude binary stars \citep{2020AJ....159..280B}.
    \item Not a giant star. We excluded giant stars using the criterion $\log_{10} {\frac{\mathrm{R_\star}}{\mathrm{R_\odot}}} < 0.00035 \times (\mathrm{T_{\rm eff}} - 4500) + 0.15$ as outlined in \cite{2017AJ....154..109F} in order to focus our study on main-sequence systems. 
    \item Planet orbital period $(\rm P)$. We excluded planet candidates with orbital periods shorter than one day, as ultra-short-period planets represent a distinct planetary population \citep{2018ApJ...864L..38D}. 
    \item Planet radius $(\rm R_{\rm p})$. We excluded giant planets (i.e., planets with radii greater than 4 $\rm R_\oplus$) and primarily focused on small-radius planets. 
    \item Detection efficiency. We considered planets with a signal-to-noise ratio (S/N) greater than 7.1, as this was the detection threshold calculated by the $Kepler$ team \citep{2011ApJ...736...19B}. 
\end{enumerate}

After applying the above selection criteria, we were left with 587 stars hosting a total of 829 planets. We categorized these planetary systems into multiple-transiting planet systems (hereafter multis) and single-transiting planet systems (hereafter singles), with 166 and 421 host stars, respectively. However, this study focused exclusively on multis. Figure \ref{figdata2HRandPR} shows the distribution of stars and planets analyzed in this paper. 

\re{During the selection criteria process, we considered stellar multiplicity and the host star’s evolutionary stage as potential factors that could influence the inner edge. To obtain a cleaner sample, we aimed to isolate the effect of stellar mass as much as possible while minimizing the impact of other factors. The formation and evolution of planetary systems around binary stars may differ significantly from those around single stars \citep[e.g.,][]{2013ARA&A..51..269D}. Binaries can alter the structure of the protoplanetary disk and affect planetary orbits \citep{2016AJ....152....8K,2022AJ....163..207C,2022MNRAS.512..648D}, thereby influencing the position of the inner edge. Moreover, the impact of the host star’s evolutionary stage on planetary orbits is more direct. As a star evolves, its temperature changes, and when it transitions into a red giant, these changes may lead to the engulfment of the innermost planets.}

\begin{figure*}[htbp]
\setlength{\abovecaptionskip}{-0.1em}
\centering
\begin{minipage}[b]{0.33\textwidth}
    \centering
    \includegraphics[width=\textwidth]{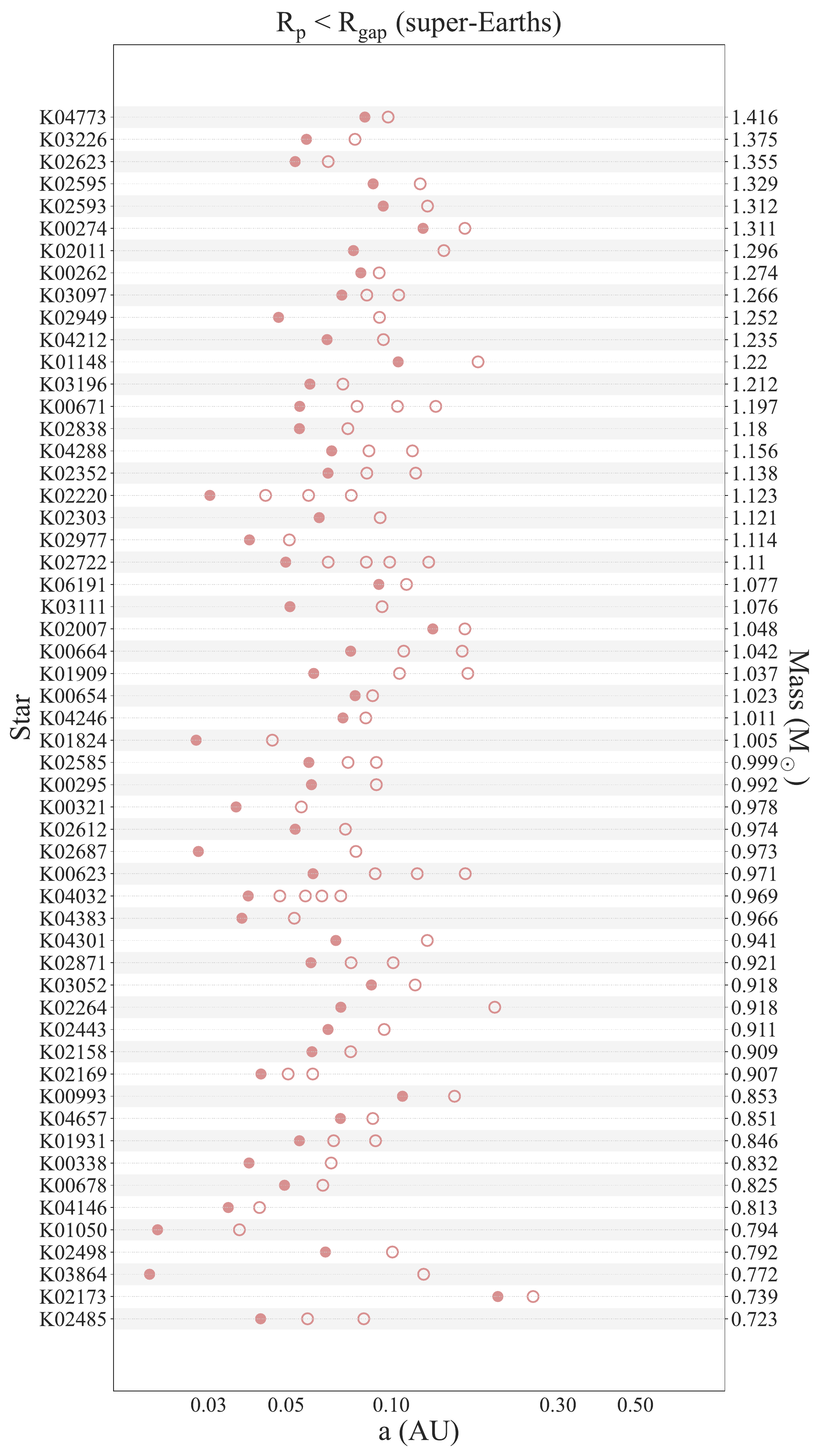}
\end{minipage}
\hfill
\begin{minipage}[b]{0.33\textwidth}
    \centering
    \includegraphics[width=\textwidth]{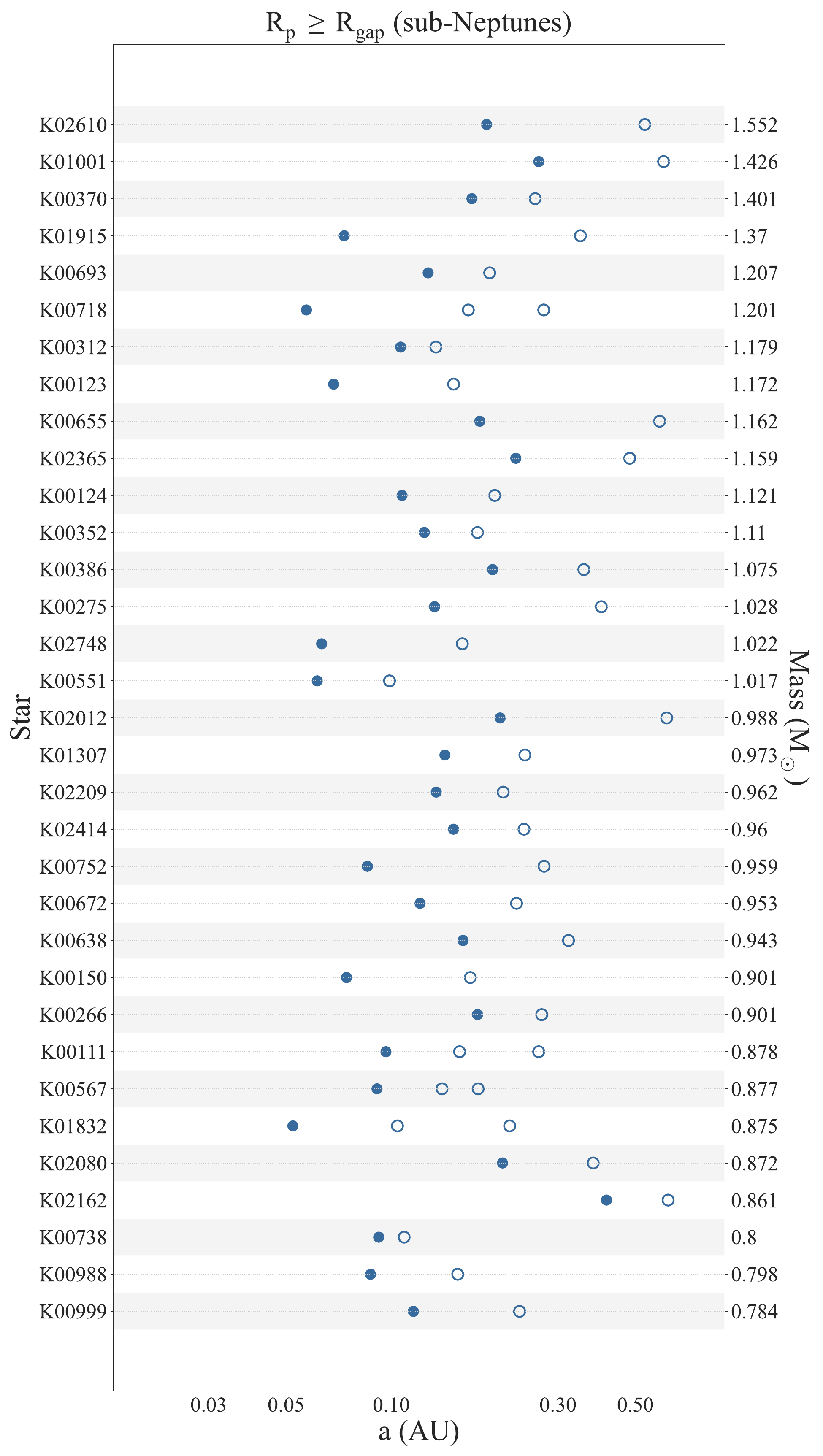}
\end{minipage}
\hfill
\begin{minipage}[b]{0.33\textwidth}
    \centering
    \includegraphics[width=\textwidth]{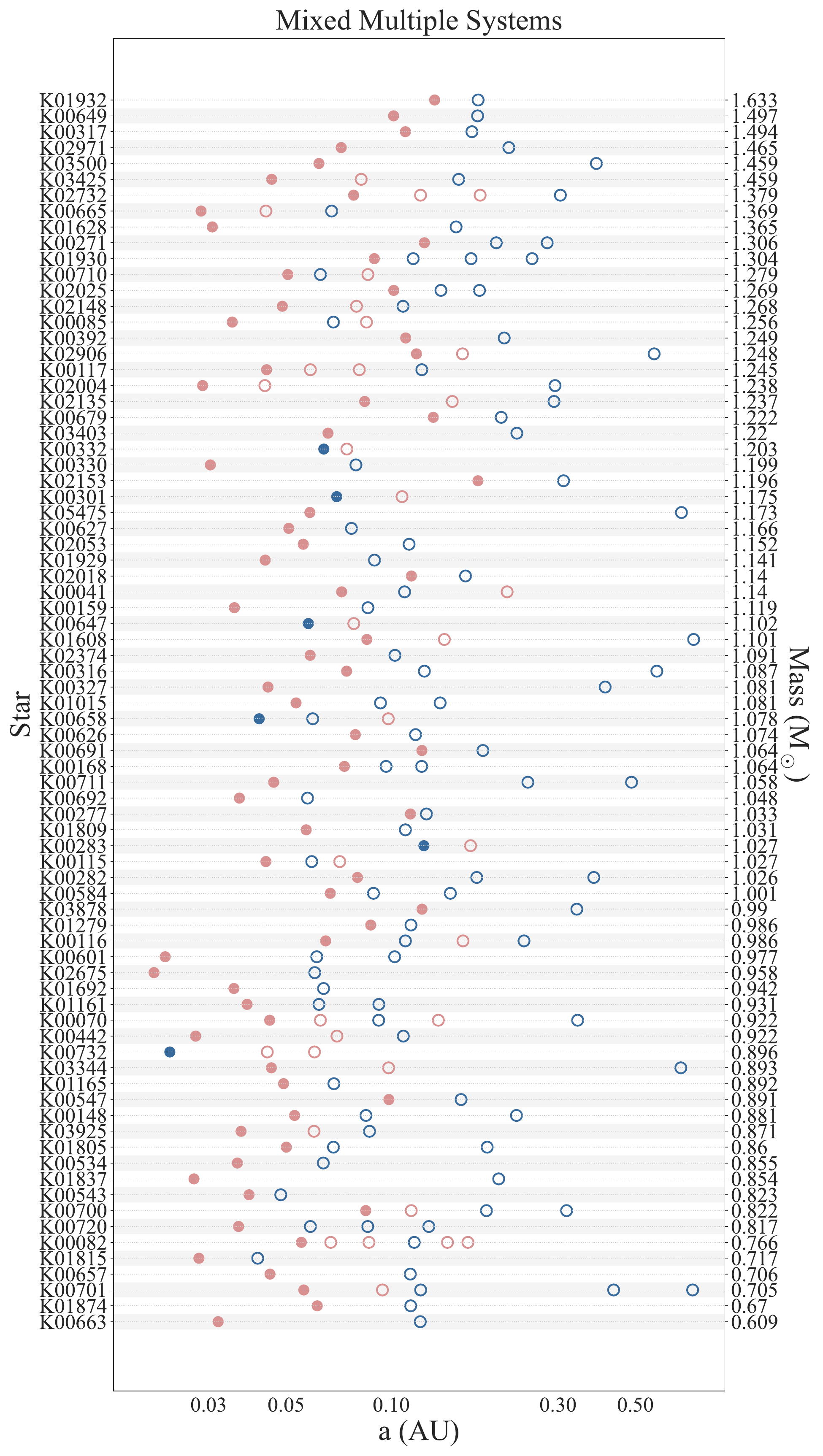}
\end{minipage}
\vspace{-1.6em}
\caption{Architecture of the three population samples of all multis studied in this paper. The left panel represents super-Earths, the middle panel represents sub-Neptunes, and the right panel represents mixed systems containing both super-Earths and sub-Neptunes within the same system. The point colors and styles are the same as those in Fig. \ref{figdata2HRandPR}. Each row represents a system, ordered from top to bottom by decreasing stellar mass. A trend can be observed: As stellar mass increases from low to high, the semimajor axis shifts from smaller to larger values.}
\label{figKBPin}
\end{figure*}

In $Kepler$ systems, singles constitute the vast majority. Due to the fact that singles tend to have larger inclinations \citep{2016PNAS..11311431X,2018ApJ...860..101Z}, the innermost planet (the true inner edge) may be missed, resulting in the observation of only the outer planets (a false inner edge), which introduces observational bias in measuring the inner edge. In contrast, multis in $Kepler$ systems are generally considered to be coplanar \citep{2014ApJ...790..146F,2016PNAS..11311431X}, making it unlikely to detect outer planets without also observing the innermost ones. However, this does not apply to ultra-short-period planets \citep{2018ApJ...864L..38D}, which have already been excluded from our sample. Therefore, we separated singles and multis, conducting independent studies on the latter. We believe that the observational results from multis are more representative of the intrinsic relationship between the inner edge and stellar properties.

\subsection{Systems division}
\label{sec:sysdivision}

Previous studies have revealed a valley around 1.9 R$_\oplus$ in the radius distribution of small planets, which separates super-Earths and sub-Neptunes \citep{2013ApJ...775..105O,2017AJ....154..109F}. \re{In this work, we used the radius valley for classification because super-Earths and sub-Neptunes may have different formation mechanisms and formation locations \citep[e.g.,][]{2024A&A...687A..25C}.} To facilitate a more detailed analysis, we calculated the radius valley using
\begin{equation}
R_{\text{gap}} = 1.9 R_\oplus \left(\frac{P}{10 \, \text{days}} \right)^{-0.09} \left(\frac{M_\star}{M_\odot} \right)^{0.26}, 
\label{radiivalley}
\end{equation}
which was derived from Eq. (12) in \cite{2021ARA&A..59..291Z}. Based on this, we divided all planetary systems in our sample into three categories: systems containing only super-Earths (R$_{\mathrm{p}}$ $<$ R$_{\mathrm{gap}}$), systems containing only sub-Neptunes (R$_{\mathrm{p}}$ $\geq$ R$_{\mathrm{gap}}$), and mixed systems that include both super-Earths and sub-Neptunes within the same system. These three categories represent completely distinct populations with no overlap.

Figure \ref{figKBPin} illustrates the architecture of the three population samples of all multis analyzed in this study. The left panel represents super-Earth systems, the middle panel represents sub-Neptune systems, and the right panel represents mixed systems containing both super-Earths and sub-Neptunes within the same system. Each panel is sorted by stellar mass. A trend is evident: As stellar mass increases from low to high, the semimajor axis shifts progressively from smaller to larger values.

In the final sample, we used the following datasets: all multiple systems, consisting of 166 systems with 408 planets; super-Earth systems, including 55 systems with 132 planets; sub-Neptune systems, containing 33 systems with 70 planets; and mixed multiple systems, comprising 78 systems with 206 planets. The “all multiple systems” dataset includes the total data from the other three populations. 

\section{Analyses and results} 
\label{sec:anaandres}

\subsection{Dependence of inner edge on stellar mass}
\label{sec:observedresults}

\begin{figure*}[htbp]
\setlength{\abovecaptionskip}{-0.2em}
\centering
\begin{minipage}[b]{0.45\textwidth}
    \centering
    \includegraphics[width=\textwidth]{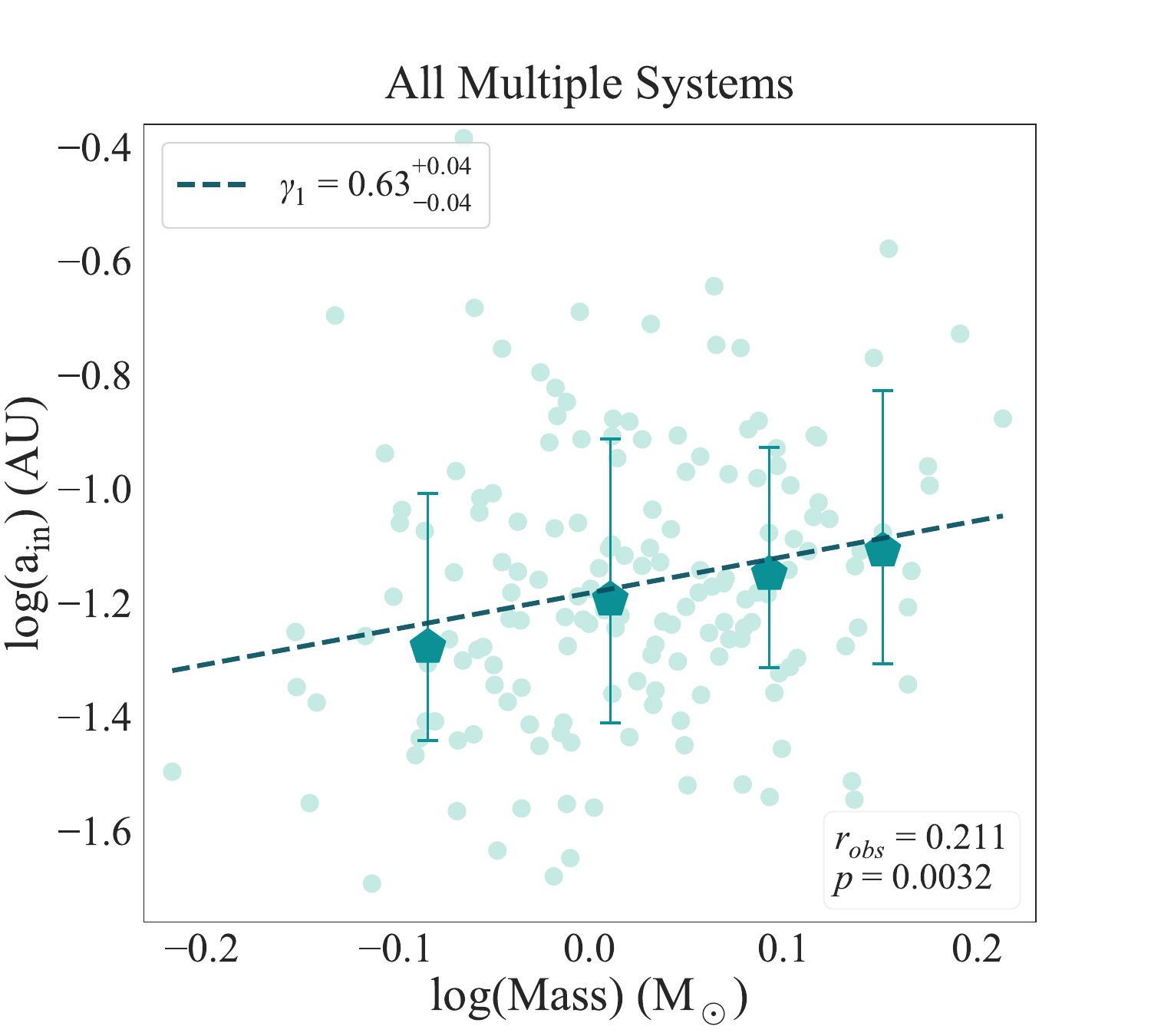}
\end{minipage}
\begin{minipage}[b]{0.45\textwidth}
    \centering
    \includegraphics[width=\textwidth]{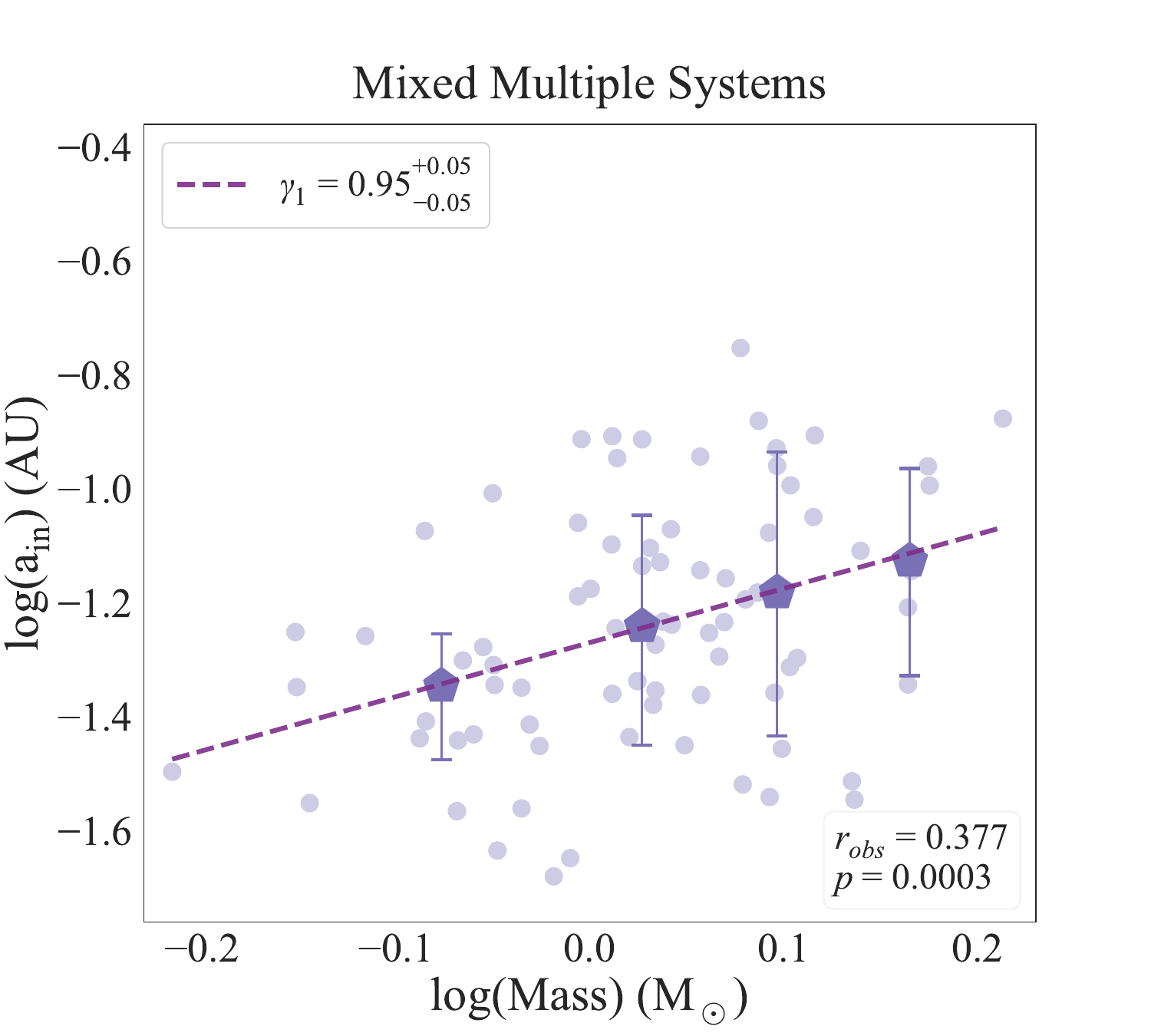}
\end{minipage}
\begin{minipage}[b]{0.45\textwidth}
    \centering
    \includegraphics[width=\textwidth]{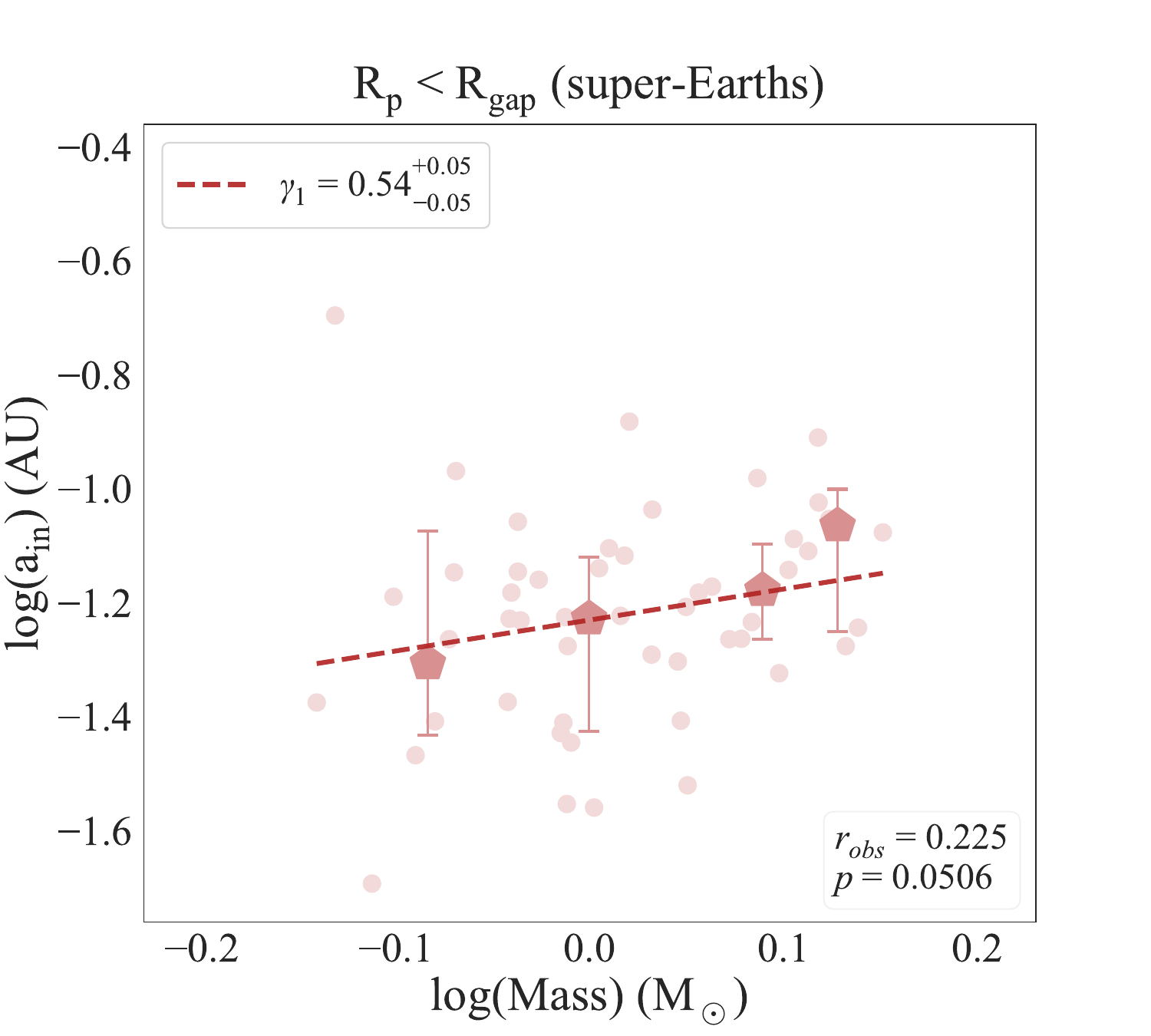}
\end{minipage}
\begin{minipage}[b]{0.45\textwidth}
    \centering
    \includegraphics[width=\textwidth]{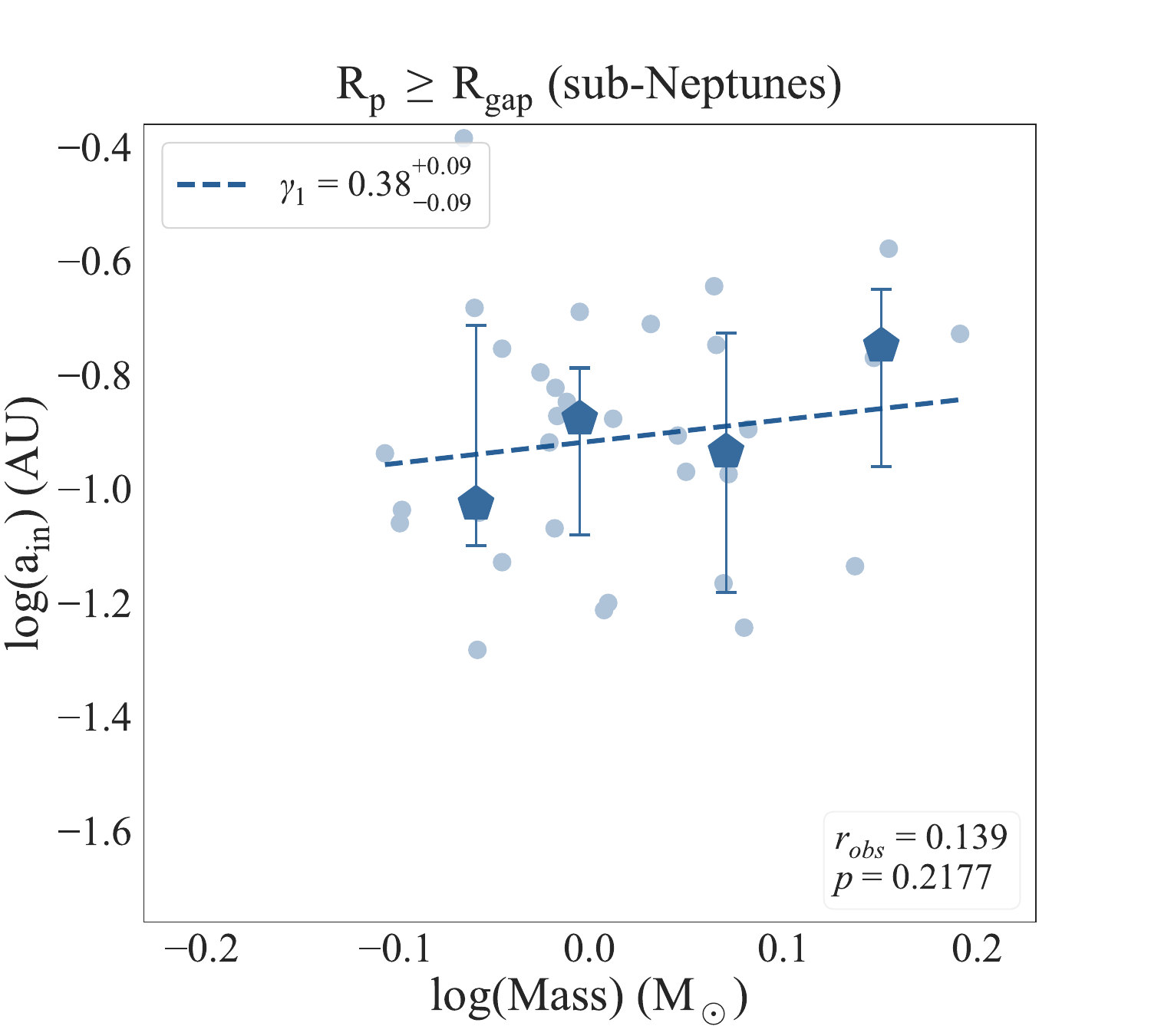}
\end{minipage}
\caption{The inner edge of planetary systems as a function of stellar mass for the datasets of all multiple systems, super-Earth systems, sub-Neptune systems, and mixed multiple systems. The four panels illustrate the correlation between the inner edge and stellar mass for different populations. Different colors are used to represent each population: green for all multiple systems, red for super-Earth systems, blue for sub-Neptune systems, and purple for mixed multiple systems. Points represent the observational data, while pentagons indicate the median value for each bin, with error bars showing $1\sigma$. In this work, the binning is merely intended to guide the eye. Dashed lines represent the fit to the observational data based on Eq. (\ref{logaM}). The power-law index corresponds to $\gamma_1$ in Eqs. (\ref{aM}) and (\ref{logaM}), with error bars also indicating $1\sigma$. $r_{obs}$ and $p$ represent the Pearson coefficients and p-values for the four populations, respectively. \re{Note: The median values and $1\sigma$ dispersion of the binned data in this study are calculated in logarithmic space and are based on actual observational data.}}
\label{figdataset1correlation}
\end{figure*}

\re{In this study, we primarily focused on the relationship between stellar mass and the inner edge. For stellar effective temperature \citep{2013ApJS..208....9P}, we selected main-sequence stars, where temperature and mass are approximately equivalent. An empirical relation exists between them, given by $T_{\text{eff}} \propto M_{\star}^{0.5}$. Therefore, by studying stellar mass, we also indirectly account for stellar temperature. Stellar age is another potentially important factor; however, we did not include it in this work due to the large uncertainties in current age estimates \citep{2020AJ....159..280B}.}

We defined the inner edge of the planetary system as the location of the innermost planet within the system. Although the orbital period and the semimajor axis are equivalent measures of orbital size, we chose to work with the semimajor axis to study the dependence of the inner edge on stellar mass for better comparison with previous observational and theoretical studies. Thus, the relationship between the inner edge ($a_{\text{in}}$) of planetary systems and stellar mass ($M_{\star}$) can be described by the following simple power-law form: 
\begin{equation}
a_{\text{in}} = \re{\gamma_{0,am}} M_{\star}^{\gamma_1},
\label{aM}
\end{equation}
and it can be converted to log space for analysis:
\begin{equation}
\log(a_{\text{in}}) = \re{\log(\gamma_{0,am})} + \gamma_1 \log(M_{\star}).
\label{logaM}
\end{equation}
\re{It is important to note that in this study, all $\gamma_{0}$ terms refer to the intercept, with different subscript letters indicating the specific combination of parameters included in each equation.} We can calculate $a_{\text{in}}$ by obtaining the planet orbital period (P) from the $Kepler$ catalog using the following formula: 
\begin{equation}
a = \sqrt[3]{\frac{GM_{\star} P^2}{4\pi^2}}. 
\label{ain}
\end{equation}

\begin{figure*}[htbp]
\setlength{\abovecaptionskip}{-2.3em}
\centering
\includegraphics[width=0.9\textwidth]{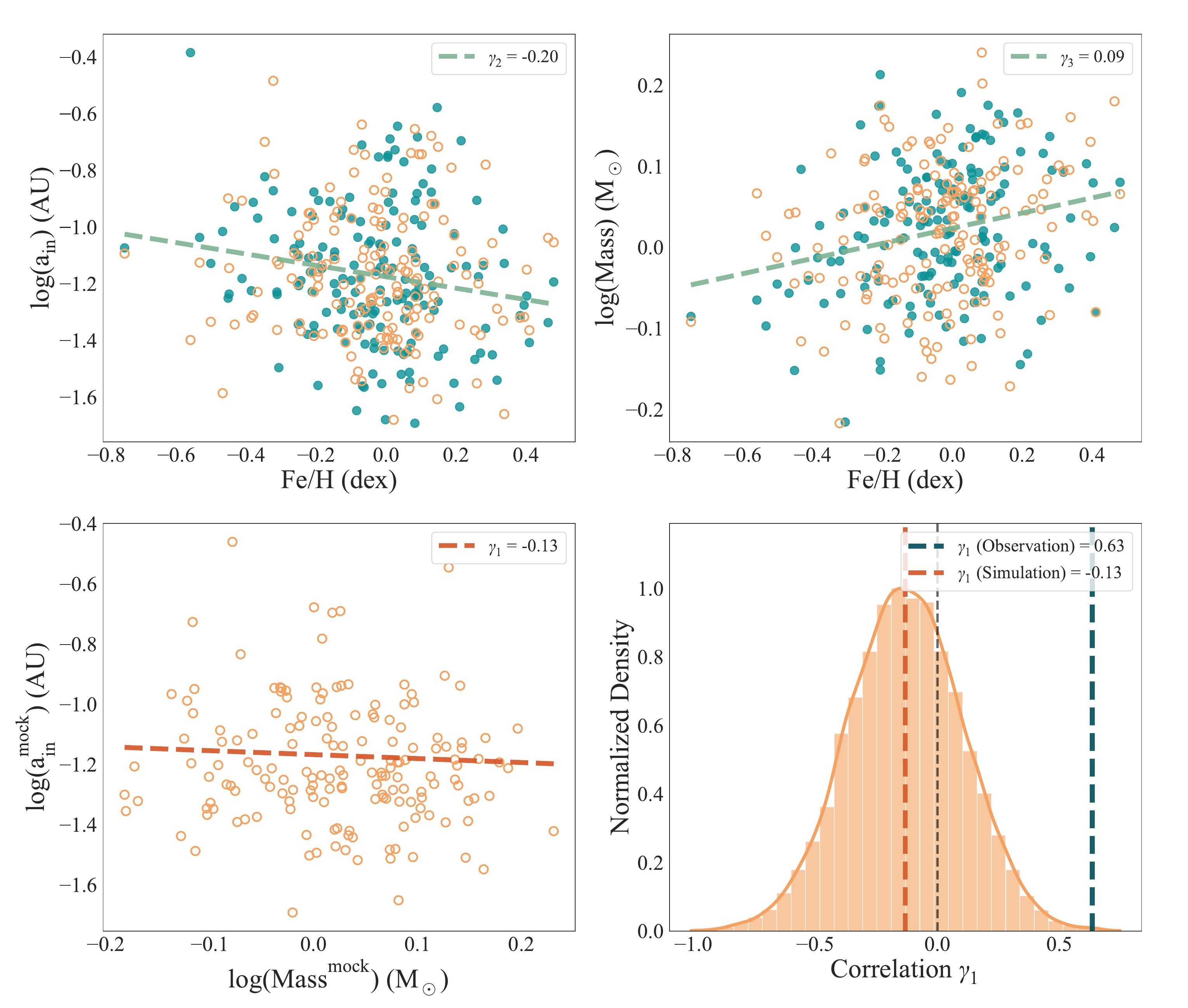}
\vspace{-0.6em}
\caption{Effect of stellar metallicity on the stellar mass\text{--}inner edge correlation. The metallicity dependence of the inner edge and stellar mass, as well as their projection (top two panels) on the inner edge and stellar mass diagram (bottom-left panel). Green solid points represent all observational samples, while orange hollow points represent mock datasets generated from the simulation process. The light green dashed line represents the fitting results for the green solid points ($\gamma_2$ and $\gamma_3$ in Eqs. (\ref{fehain}) and (\ref{fehmass})), while the orange dashed line represents the fitting results for the orange hollow points ($\gamma_1$ in Eq. (\ref{logaM})). The top two panels and the bottom-left panel collectively display the median result from the 10,000 simulation runs. The bottom-right panel shows the distribution of the correlation index ($\gamma_1$) between the inner edge and stellar mass for the mock data (using Kernel Density Estimation), with the orange dashed line indicating the median of the distribution, the black dashed line indicating a correlation of zero, and the green dashed line representing the correlation of all observational data (as shown in the top-left panel of Fig. \ref{figdataset1correlation}). Note: To further improve our analysis of stellar metallicity, we removed two outliers (with values less than $-1.0$) from the correlation analysis. 
\label{figdata2fehcorrelationmul}}
\end{figure*}

Using Eq. (\ref{logaM}) and applying the least squares method for fitting, we presented the observational results of this relationship. From Fig. \ref{figdataset1correlation}, a correlation between stellar mass and the inner edge can be observed across different planetary populations, with the strength of this correlation ($\gamma_1$) varying among them. The four panels represent the four populations mentioned in Sect. \ref{sec:sysdivision}. For all multiple systems, super-Earths, sub-Neptunes, and mixed multiple systems, the correlation index $\gamma_1$ is \re{$0.63_{-0.04}^{+0.04}$, $0.54_{-0.05}^{+0.05}$, $0.38_{-0.09}^{+0.09}$, and $0.95_{-0.05}^{+0.05}$,} respectively. The uncertainties of these results were calculated by resampling stellar mass within the error range provided in Berger20, recalculating $a_{\text{in}}$ using Eq. (\ref{ain}), and repeating the fitting process 10,000 times. All error bars in this paper represent the $1\sigma$ range.

To further investigate the correlation between stellar mass and the inner edge, we performed a Pearson coefficient test to quantify this correlation. The results are displayed in the bottom-right corner of each panel (Fig. \ref{figdataset1correlation}). It can be observed that all populations exhibit correlations, but the strength and confidence level of these correlations vary. For all multiple systems, super-Earths, sub-Neptunes, and mixed multiple systems, the Pearson coefficients ($r_{obs}$) are 0.211, 0.225, 0.139, and 0.377, respectively. It is worth remarking at this point that these are relatively modest correlations, and we discuss the scatter in Sect. \ref{futurework}. To calculate the p-value, we shuffled the observed data and randomly paired stellar masses and inner edges to create a new dataset matching the size of the observed data. We then calculated the Pearson coefficient ($r$) for this random dataset. Repeating this process 100,000 times ($N_{tot}$), we compared the $r$ values with $r_{obs}$ and counted the number of cases where $r > r_{obs}$, denoted as $N(r > r_{obs})$. Thus, the p-value was calculated as $N(r > r_{obs}) / N_{tot}$. A smaller p-value indicates a more reliable correlation. The results show that, for all multiple systems, super-Earths, sub-Neptunes, and mixed multiple systems, the p-values ($p$) are \re{0.0032, 0.0506, 0.2177, and 0.0003,} respectively. Therefore, the correlations for all multiple systems and mixed multiple systems are stronger and more significant, while the correlations for super-Earths and sub-Neptunes are weaker, especially for sub-Neptunes. The weaker confidence levels for these two populations may be due to smaller sample sizes. \re{Here, we performed simulations to test the quantitative impact of sample size on the results for super-Earths and sub-Neptunes. This was tested by randomly drawing the same number of planets from each population within the all multiple systems sample and analyzing the resulting distribution of slope values over 10,000 simulations. The results show that the median and $1\sigma$ range of the slope distributions for super-Earths and sub-Neptunes are $0.63_{-0.35}^{+0.35}$ and $0.65_{-0.49}^{+0.49}$, respectively. It is clear that both distributions have large error bars, and the observed slopes of $0.54_{-0.05}^{+0.05}$ and $0.38_{-0.09}^{+0.09}$ fall within the $1\sigma$ range of the simulated distributions. Therefore, we cannot rule out the possibility that the observed difference between super-Earths and sub-Neptunes is due to limited sample sizes.} 

We note that, at this point, we have not considered the influence of other stellar properties. In the following sections, we further investigate whether this apparent correlation is influenced by other factors.

\subsection{Effect of stellar metallicity}
\label{sec:effectofageandfeh}

\re{Previous studies have demonstrated that stellar metallicity influences the architecture of planetary systems. \cite{2013ApJ...763...12B} reported, based on statistical analysis, that stellar metallicity affects the distribution of planetary periods, which could, in turn, influence the position of the inner edge. Similarly, \cite{2016AJ....152..187M} found that short-period planets—particularly hot rocky planets—are more likely to orbit host stars with higher metallicity.}

Since stellar metallicity influences planetary architecture (e.g., orbital periods) and is also correlated with stellar mass, the effect of metallicity should be considered in order to remove its impact on the relationship between the inner edge and stellar mass. We plot the top two panels of Fig. \ref{figdata2fehcorrelationmul}, showing the inner edge as a function of metallicity and stellar mass as a function of metallicity, respectively. As can be seen from the fitting results of the observed data (represented by the green solid points and the light green dashed lines), the inner edge exhibits a negative correlation with metallicity \re{($\gamma_2 = -0.20$)}, while metallicity shows a positive correlation with stellar mass \re{($\gamma_3 = 0.09$)}. To investigate the projection of these two relationships on the inner edge\text{--}stellar mass diagram, we conducted the following analysis of data mocking and fitting.

We generated the mock data of the inner edge and stellar mass with the specific calculations corresponding to the following functional forms:
\begin{equation}
\log(a_{\text{in}}^{\text{mock}}) = \re{\log(\gamma_{0,afr})} + \gamma_2 \, \text{[Fe/H]} + \text{res}(a_{\text{in}})
\label{fehain}
\end{equation}
and 
\begin{equation}
\log(M_{\star}^{\text{mock}}) = \re{\log(\gamma_{0,mfr})} + \gamma_3 \, \text{[Fe/H]} + \text{res}(M_{\star}), 
\label{fehmass}
\end{equation}
where \re{$\gamma_{0,afr}$} and $\gamma_2$ in Eq. (\ref{fehain}) correspond to the best fit in the top-left panel of Fig. \ref{figdata2fehcorrelationmul}, and \re{$\gamma_{0,mfr}$} and $\gamma_3$ in Eq. (\ref{fehmass}) correspond to the best fit in the top-right panel of Fig. \ref{figdata2fehcorrelationmul}. The terms $\text{res}(a_{\text{in}})$ and $\text{res}(M_{\star})$ are the residuals, which are randomly selected from the residuals of the top-left panel and top-right panel, respectively. In the bottom-left panel of Fig. \ref{figdata2fehcorrelationmul}, we show a typical case of data mocking and fitting. Fitting this mock data (orange hollow points) using Eq. (\ref{logaM}) produced an anticorrelation on the $a_{\text{in}} - M_{\star}$ diagram (represented by the orange dashed line with a $\gamma_1$ power-law index value of \re{$-0.13$}).

The above data mocking and fitting process was repeated 10,000 times, and the distribution of the correlation index ($\gamma_1$) between the inner edge and stellar mass for the mock data is illustrated in the bottom-right panel of Fig. \ref{figdata2fehcorrelationmul}. As can be seen, the median power-law index of the correlation distribution is \re{$-0.13$}. In contrast, our observational results show a positive correlation between the inner edge and stellar mass of \re{0.63} (the correlation for all multiple systems, as shown in the top-left panel of Fig. \ref{figdataset1correlation}), which is significantly larger than the median of the distribution. Furthermore, out of 10,000 data mocking and fitting cases, only six produced a positive correlation with \re{$\gamma_1 > 0.63$}, corresponding to a p-value of 0.0006 for a hypothesis that the observed positive correlation was caused by the projection of the above metallicity effect (Eqs. (\ref{fehain}) and (\ref{fehmass})). 

Therefore, it is necessary to separately consider the dependence of the inner edge on both stellar mass and metallicity. If the negative impact of the metallicity-induced projection is corrected, the correlation between $a_{\text{in}}$ and $M_{\star}$ should be even stronger. In addition, since different populations of planets are affected by metallicity in varying ways, the degree of correction to the correlation between the inner edge and stellar mass may also vary among different planetary populations.

\subsection{Multiple linear regression model}
\label{sec:MLR}

\begin{figure*}[htbp]
\centering
\includegraphics[width=0.9\textwidth]{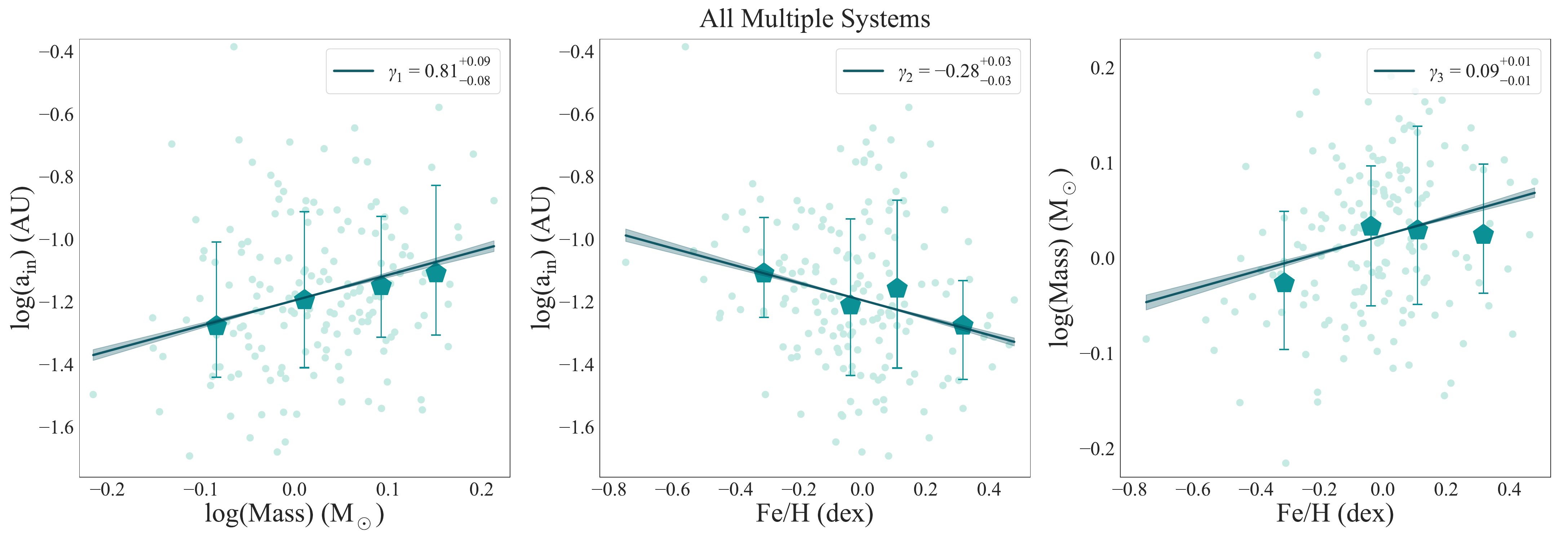}
\vspace{-0.4em}
\includegraphics[width=0.9\textwidth]{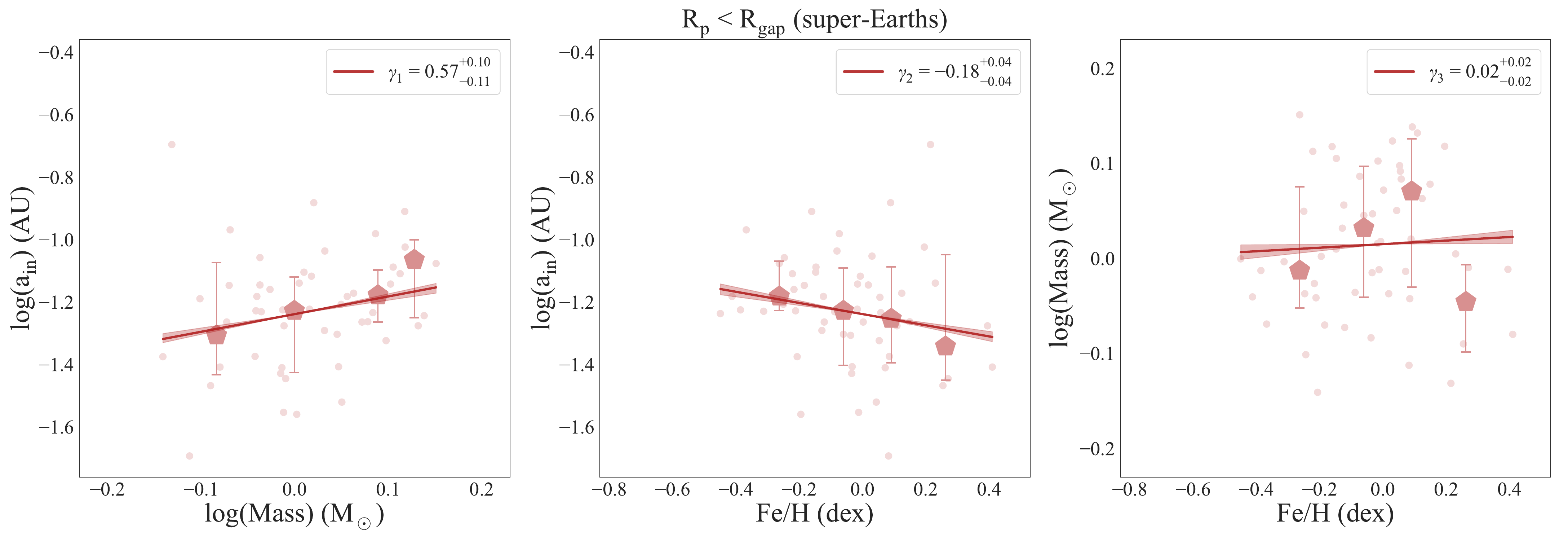}
\vspace{-0.4em}
\includegraphics[width=0.9\textwidth]{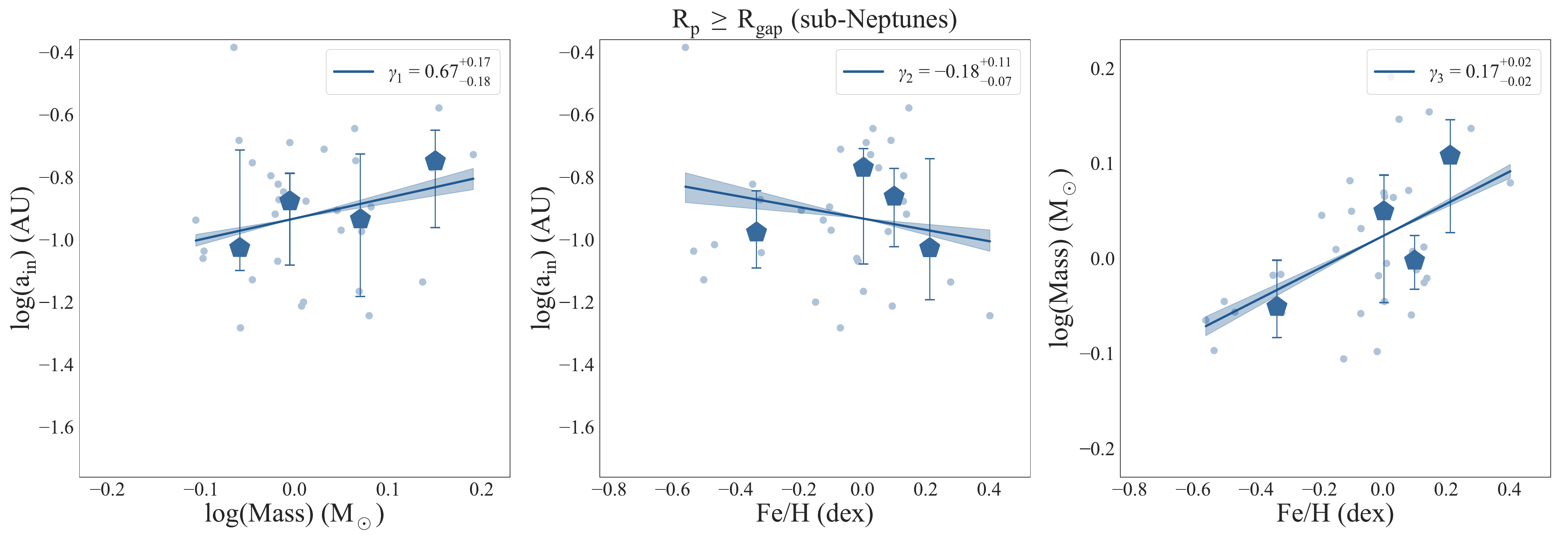}
\vspace{-0.4em}
\includegraphics[width=0.9\textwidth]{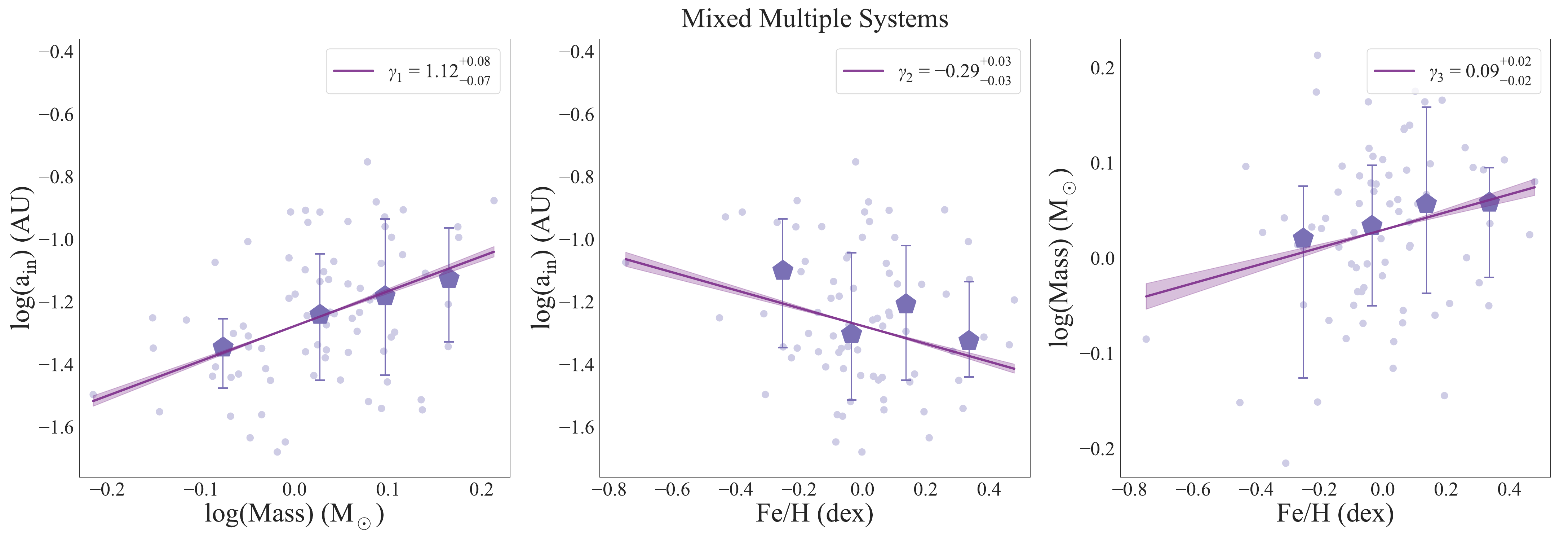}
\vspace{-0.4em}
\caption{Dependence of the inner edge on stellar mass and metallicity as well as the corresponding dependence between mass and metallicity obtained through the MLR model. The left and middle panels show the power-law indices, which correspond to $\gamma_1$ and $\gamma_2$ in Eqs. (\ref{lrall}) and (\ref{lrlogall}), respectively. Additionally, the right panels show the power-law index corresponding to $\gamma_3$ in Eq. (\ref{fehlog}). The lines represent the fitting results, with color bands indicating $1\sigma$ error bars. The colors and styles of points and lines are consistent with those in Fig. \ref{figdataset1correlation}. \re{Note: The median values and $1\sigma$ dispersion of the binned data in this study are calculated in logarithmic space and are based on actual observational data.}}
\label{figml}
\end{figure*}

To investigate the dependence of the inner edge on both stellar mass and metallicity, we utilized the multiple linear regression (MLR) model. The MLR identifies the best-fitting straight line to describe the relationship between multiple input variables and the output variable. Although the relationship between stellar mass, metallicity, and the inner edge was not strictly linear, we converted it into a multiple linear relationship through logarithmic transformation and analyzed it using an MLR model. To better express the relationships between stellar mass, metallicity, and the inner edge, we adopted the following relational equation, a simple power-law model in the form: 
\begin{equation}
\frac{a_{\text{in}}}{\text{AU}} = \re{\gamma_{0,amf}} \left(\frac{M_{\star}}{M_{\odot}} \right)^{\gamma_1} 10^{\gamma_2 \, \text{[Fe/H]}}.
\label{lrall}
\end{equation}
The equation became easier to work with in logarithmic space:
\begin{equation}
\log\left(\frac{a_{\text{in}}}{\text{AU}} \right) = \re{\log(\gamma_{0,amf})} + \gamma_1 \log\left(\frac{M_{\star}}{M_{\odot}} \right) + \gamma_2 \, \text{[Fe/H]}.
\label{lrlogall}
\end{equation} 
\re{Equation (\ref{lrall}) mainly follows previous studies on occurrence rates, such as Eq. (1) in \cite{2010PASP..122..905J} and Eq. (21) in \cite{2021ARA&A..59..291Z}. However, these formulas do not have a strong theoretical or astrophysical basis. Although they are not specifically designed for studying the inner edge, we believe they can serve as a starting point for both preliminary and detailed analyses.} \ree{More importantly, using these formula forms allows us to directly compare our results with previous observational findings and theoretical models (see Sect. \ref{sec:impanddis}), thereby making the results more meaningful for interpretation and discussion.}

Using the MLR model from Eq. (\ref{lrlogall}), we derived the dependence of the inner edge on stellar mass and metallicity for each population dataset. \re{We adopted the same method as in Sect. \ref{sec:observedresults} to report the uncertainties for the slopes, which were derived by accounting for individual data point uncertainties and repeating the process 10,000 times.} In Fig. \ref{figml}, the four rows of panels present the results of the four populations obtained through the MLR model. The left panels show the distribution of the inner edge and stellar mass, as well as the fitting results for all multiple systems, super-Earths, sub-Neptunes, and mixed multiple systems, with $\gamma_1$ values of \re{$0.81_{-0.08}^{+0.09}$, $0.57_{-0.11}^{+0.10}$, $0.67_{-0.18}^{+0.17}$, and $1.12_{-0.07}^{+0.08}$}, respectively. The middle panels display the distribution of the inner edge and metallicity, as well as the fitting results, with $\gamma_2$ values of \re{$-0.28_{-0.03}^{+0.03}$, $-0.18_{-0.04}^{+0.04}$, $-0.18_{-0.07}^{+0.11}$, and $-0.29_{-0.03}^{+0.03}$}, respectively. Additionally, we determined the corresponding relationship between stellar mass and metallicity by fitting it using the least squares method through the following equation:
\begin{equation}
\log(M_{\star}) = \re{\log(\gamma_{0,mf})} + \gamma_3 \, \text{[Fe/H]}.
\label{fehlog}
\end{equation}
The right panels in Fig. \ref{figml} show the distribution of stellar mass and metallicity, as well as the fitting results, with $\gamma_3$ values of \re{$0.09_{-0.01}^{+0.01}$, $0.02_{-0.02}^{+0.02}$, $0.17_{-0.02}^{+0.02}$, and $0.09_{-0.02}^{+0.02}$}, respectively.

According to Fig. \ref{figml}, we observe that the inner edge shows a positive correlation with mass and a negative correlation with metallicity. From all multiple systems and mixed multiple systems populations, we observe that, under similar influences of metallicity on the inner edge (middle panels showing $\gamma_2$ values of approximately \re{$-0.28$ and $-0.29$}, respectively), the correlation between the inner edge and stellar mass increases by a similar magnitude after correcting for the metallicity effect (left panels compared to the results in Fig. \ref{figdataset1correlation}, $\gamma_1$ values of approximately 0.2). However, in super-Earths and sub-Neptunes populations, under similar influences of metallicity on the inner edge (middle panels showing $\gamma_2$ values of approximately \re{$-0.18$}), the correlation between the inner edge and stellar mass shows varying degrees of increase after correcting for the metallicity effect. For super-Earths, the increase is minimal (left panel compared to the result in Fig. \ref{figdataset1correlation}, $\gamma_1$ value of approximately \re{0.03}), whereas for sub-Neptunes, the increase is more pronounced (left panel compared to the result in Fig. \ref{figdataset1correlation}, $\gamma_1$ value of approximately 0.3). We believe this difference is primarily due to the correlation between stellar mass and metallicity ($\gamma_3$ in Eq. (\ref{fehlog})). The values of $\gamma_3$ for all multiple systems and mixed multiple systems are similar (right panels showing $\gamma_3$ values of approximately \re{0.09}), but there is a significant difference between super-Earths and sub-Neptunes (right panels showing $\gamma_3$ values of approximately \re{0.02 and 0.17}, nearly an order of magnitude). Sub-Neptunes have the largest $\gamma_3$, while super-Earths have the smallest $\gamma_3$. Therefore, the middle and right panels together determine the degree of metallicity correction, and the strength of the correction effect varies across the four populations. This may explain the varying degrees of increase (the change in $\gamma_1$ from Fig. \ref{figdataset1correlation} to Fig. \ref{figml}) observed among the four populations.

In summary, after accounting for the effects of stellar mass and metallicity, the dependence of the inner edge on stellar mass increases across all populations (Fig. \ref{figml} compared to Fig. \ref{figdataset1correlation}), reaching $\gamma_1$ values of approximately $0.6-1.1$. This comparison can also be observed in Table \ref{tab:powerlawindex}, which highlights the differences between the observational results. This suggests that the influence of metallicity weakens the observed dependence of the inner edge on stellar mass, indicating that the intrinsic dependence is likely stronger.

\ree{We note that modifying the functional form may lead to different strengths of the positive correlation between the inner edge and stellar mass, as well as the negative correlation between the inner edge and stellar metallicity. After testing various alternatives, we conclude that caution should be exercised when interpreting the exact numerical values. However, given the current state of research, the functional form adopted in this study is considered reasonable, as it helps to explain the observed trends and facilitates direct comparison with previous observational studies and theoretical models (see Sect. \ref{sec:impanddis}), thereby leading to more meaningful results and conclusions.}

\subsection{Effect of observational bias}
\label{sec:effectoobsbias}

For our observational study, a crucial step was to assess whether observational selection bias could potentially influence our sample analysis results. Our planet samples are from the $Kepler$ survey, the major observational bias stems from the transit geometric selection effect and the detection efficiency of the $Kepler$ telescope. Since the detection efficiency does not vary significantly with stellar temperature or mass \citep[as shown in Fig. 11 of][]{2020AJ....159..164Y}, we believe it will not significantly affect the correlation between the inner edge and stellar mass. Thus, in this section, we focus on the effect of transit geometric selection bias on the correlation between the inner edge and stellar mass. Specifically, assuming that the inner edge and stellar mass are uncorrelated, we performed simulations to model the transit selection process and determined whether any correlation or lack thereof is induced.

\begin{figure*}[htbp]
\centering
\includegraphics[width=\textwidth]{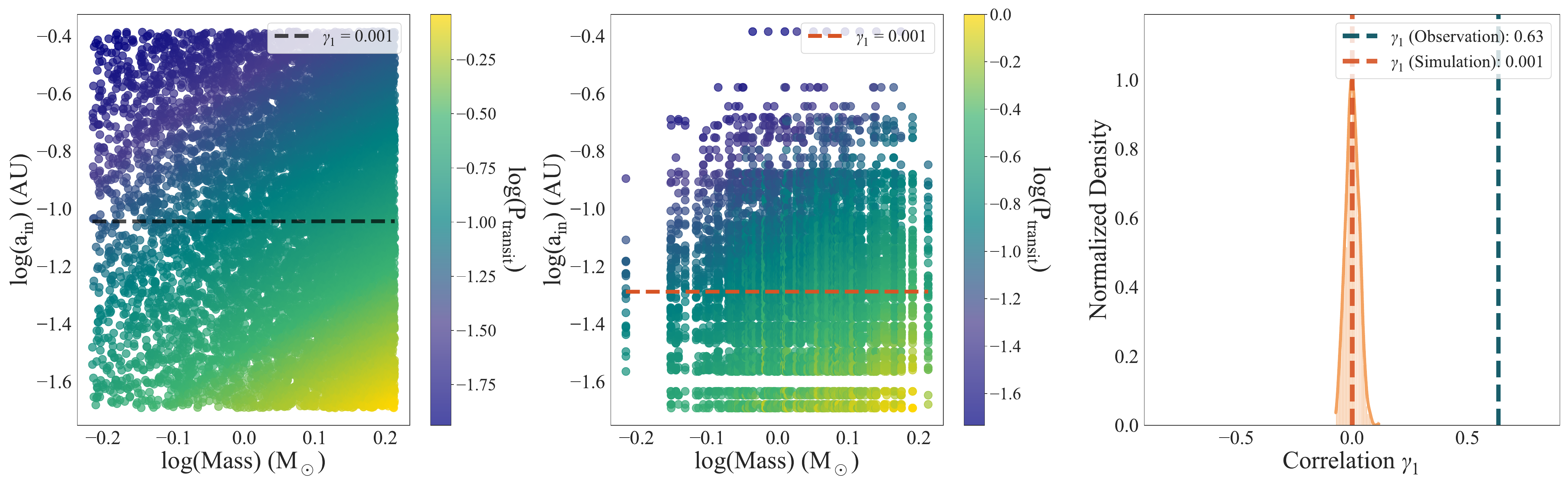}
\includegraphics[width=\textwidth]{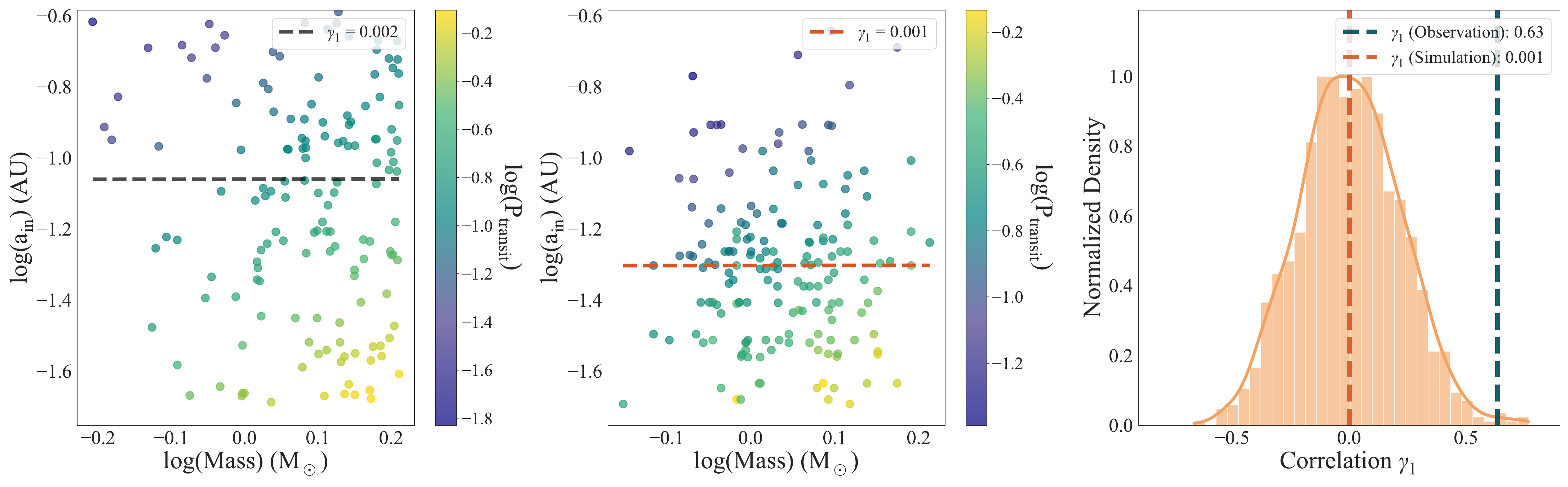}
\caption{Simulated analysis of observational selection bias. The panels in the top row represent $N_{\text{test}}$ (with a larger number of sample sets set to 10,000), while those in the bottom row represent $N_{\text{obs}}$ (the same number of sample sets as the observational data). The left panels show the results of uniformly random sampling within the range, and the middle panels show the random sampling in a way that uses observational data but applies the biases of transit detection. Points represent the distribution of simulated data, with the color bar indicating the distribution of $P_{\text{transit}}$. Black and orange dashed lines indicate the fit results ($\gamma_1$ of the correlation between the inner edge and stellar mass). Each of these four panels represents the median result from 1,000 simulation runs. The right panels display the distribution of correlations between $M_\star$ and $a_{\text{in}}$ obtained from 1,000 simulations, with meanings consistent with Fig. \ref{figdata2fehcorrelationmul}.}
\label{figdata2obsselectmul}
\end{figure*}

The specific analysis steps are as follows, and the results are shown in Fig. \ref{figdata2obsselectmul}. 
(i) A stellar mass ($M_\star$) and its corresponding stellar radius ($R_\star$) were randomly selected from the observational data, followed by a random selection of the inner edge ($a_{\text{in}}$). Unlike the observational data, we shuffled the pairing between $a_{\text{in}}$ and $M_\star$ by randomly selecting them independently from the observational data. 
(ii) The normalized transit probability was calculated using the following formula:
\begin{equation}
P_{\text{transit}} = \frac{R_\star}{a_{\text{in}}} \Big/ P_{\text{max}}, 
\label{Ptransit}
\end{equation}
where
\begin{equation}
P_{\text{max}} = \frac{R_{\star(\text{max})}}{a_{\text{in(min)}}}.
\label{Pmax}
\end{equation}
Here, $P_{\text{max}}$ represents the maximum possible transit probability in the simulation, $R_{\star(\text{max})}$ represents the largest $R_\star$ in the sample, and $a_{\text{in(min)}}$ represents the smallest $a_{\text{in}}$ in the sample. 
(iii) A random number was generated from a uniform distribution between zero and one. If $P_{\text{transit}}$ was greater than the random number, then the data generated in the first step (i.e., $M_\star$, $R_\star$, and $a_{\text{in}}$) were retained. Otherwise, we returned to the first step to regenerate another set of $M_\star$, $R_\star$, and $a_{\text{in}}$. \re{This is the core step of this section. We simulated observations based on the relative probability of transits, assigning different weights to individual data points, where the weights reflected their relative probabilities. Targets with higher probabilities—meaning they are more likely to be observed—were selected for the next stage of analysis, which focused on observational selection bias.}

This process was continued until the same number of sample sets ($N_{\text{obs}}$) matching the observational data was obtained (points as shown in the bottom-middle panel of Fig. \ref{figdata2obsselectmul}). The $N_{\text{obs}}$ sets of ($M_\star$, $R_\star$, $a_{\text{in}}$) were fit to determine the correlation $\gamma_1$ power-law index between $M_\star$ and $a_{\text{in}}$ (orange dashed line shown in the bottom-middle panel of Fig. \ref{figdata2obsselectmul}). 

According to the result in the bottom-middle panel of Fig. \ref{figdata2obsselectmul}, we present a typical case to demonstrate the effect of transit selection bias on the correlation analysis between the inner edge and stellar mass. We observe that transit observational selection bias has two effects. On the one hand, smaller stellar masses (thus smaller radii) and larger inner edges (represented by the dark points in the upper-left of the bottom-middle panel) have lower $P_{\text{transit}}$, causing transit observational selection bias to filter out these data points. This reduces the number of points in the upper-left of the bottom-middle panel and introduces a positive correlation between $M_\star$ and $a_{\text{in}}$. \re{On the other hand, regions with high data density and high $P_{\text{transit}}$ (indicated by the bright points in the lower-right of the bottom-middle panel) carry greater weight in the data fitting process. Without the effect from the lower-right corner, the fitted slope would be steeper due to observational selection bias. It is precisely this effect that leads to a flatter fitted line.} This effect counteracts the previous one and simultaneously weakens the correlation between $M_\star$ and $a_{\text{in}}$. Consequently, \re{the correlation between $M_\star$ and $a_{\text{in}}$ induced by the transit observational selection bias is minimal ($\gamma_1 = 0.001$).}

The entire process was repeated 1,000 times to produce the distribution of the correlation $\gamma_1$ power-law index for $M_\star$ and $a_{\text{in}}$ (illustrated in the bottom-right panel of Fig. \ref{figdata2obsselectmul}). As can be seen, the median power-law index of the correlation distribution is \re{0.001}. In contrast, our observational results show a positive correlation between the inner edge and stellar mass of \re{0.63} (the correlation for all multiple systems, as shown in the top-left panel of Fig. \ref{figdataset1correlation}), which is significantly larger than the median of the distribution. \re{Additionally, in 1,000 simulations, almost no positive correlation with $\gamma_1$ reaching 0.63 is produced. Therefore, we conclude that while observational selection bias may influence the slope, it is not solely responsible for the observed correlation.}

In order to check whether the above results rely on the observational data, we repeated the above test with a modification to the analysis in the first step. We provided the flat results, where the data were generated uniformly from the plane of $M_\star - a_{\text{in}}$ (points as shown in the bottom-left panel of Fig. \ref{figdata2obsselectmul}), rather than from the observational sample. In this case, $M_\star$ and $a_{\text{in}}$ are randomly selected within their respective ranges, while $R_\star$ is recalculated based on the fit between mass and radius from the observational data. The result is shown in the bottom-left panel of Fig. \ref{figdata2obsselectmul}. We present a typical case to demonstrate the effect of transit selection bias on the correlation analysis between the inner edge and mass, where $\gamma_1$ is \re{$0.002$} (black dashed line as shown in the bottom-left panel of Fig. \ref{figdata2obsselectmul}). The result is almost identical to the bottom-middle panel, indicating that the test result is robust regardless of the specific data.

To ensure a thorough verification, we repeated the above test but changed the number of sample sets ($N_{\text{test}}$ = 10,000). As shown in the top-left and middle panels of Fig. \ref{figdata2obsselectmul}, the correlation $\gamma_1$ of the relationship between stellar mass and the inner edge is close to zero (black dashed line and orange dashed line). The distribution in the top-right panel of Fig. \ref{figdata2obsselectmul} indicates that the median power-law index of the correlation distribution is 0.001, which is far from the observed positive correlation $\gamma_1$ (\re{0.63}). In addition, compared to the bottom-right panel, the distribution of $\gamma_1$ in the top-right panel is more concentrated due to the use of a larger number of sample sets (10,000) in the simulation.

To summarize the above tests, we find that the transit selection bias does not induce a significant correlation. This suggests that the observed correlation between stellar mass and the inner edge is not driven by observational bias.

\section{Discussion}
\label{sec:impanddis}

\begin{table*}[htbp]
\renewcommand\arraystretch{1.5}
\centering
\caption{\centering Power-law index of the result and model.}
\label{tab:powerlawindex}
{\footnotesize
\linespread{1.9}
\begin{tabular}{c|l|c|c} 
\hline\hline
\multicolumn{2}{c|}{Dataset and Mechanism} & Index ($\gamma_1$)\textsuperscript{1} & Reference \\ \hline
\multirow{5}{*}{Observational Result} 
& \multirow{2}{*}{{Previous Work}} & $0.38$ – $0.9$ & {\citet{2013ApJ...769...86P}} \\
& & $1/3 \sim \re{0.33}$ & {\citet{2015ApJ...798..112M}} \\ \cline{2-4}
& all multiple systems & \re{{$0.81_{-0.08}^{+0.09}$} ($0.63_{-0.04}^{+0.04}$)} & \multirow{4}{*}{{This Work}\textsuperscript{2}} \\
& R$_{\mathrm{p}}$ $<$ R$_{\mathrm{gap}}$ (super-Earths) & \re{$0.57_{-0.11}^{+0.10}$ ($0.54_{-0.05}^{+0.05}$)} & \\
& R$_{\mathrm{p}}$ $\geq$ R$_{\mathrm{gap}}$ (sub-Neptunes) & \re{$0.67_{-0.18}^{+0.17}$ ($0.38_{-0.09}^{+0.09}$)} & \\
& mixed multiple systems & \re{$1.12_{-0.07}^{+0.08}$ ($0.95_{-0.05}^{+0.05}$)} & \\ \cline{1-4}
\multirow{6}{*}{Theoretical Model} 
& Sublimation (active, ${\alpha} = 2$)\textsuperscript{3} & $11/9 \sim \re{1.22}$ & {\citet{2019A&A...632A...7L}} \\
& Sublimation (passive, ${k} = 2$)\textsuperscript{4} & {$1.0$} & {\citet{2001ApJ...560..957D}} \\
& Sublimation (active, ${\alpha} = 1$)\textsuperscript{3} & {$7/9 \sim \re{0.78}$} & {\citet{2019A&A...632A...7L}} \\
& Stellar Tides & $9/13 \sim \re{0.69}$ & {\citet{2009ApJ...698.1357J}} \\
& Sublimation (passive, ${k} = 1$)\textsuperscript{4} & {$0.5$} & {\citet{2001ApJ...560..957D}} \\
& Co-rotation Radius & $1/3 \sim \re{0.33}$ & {\citet{2015ApJ...798..112M}} \\
& Planetary Tides & $3/13 \sim \re{0.23}$ & {\citet{2009ApJ...698.1357J}} \\ \hline
\end{tabular}
}
\tablefoot{
\begin{enumerate}
    \item The index represents the power-law relationship between the inner edge and stellar mass ($\gamma_1$).
    \item To compare with the degree of metallicity correction presented in the observational results of this work, it is expressed here in the form: “$\gamma_1$ in Eq. (\ref{lrlogall}) ($\gamma_1$ in Eq. (\ref{logaM}))”.
    \item In the theoretical model of active dust sublimation, $\alpha$ represents the power-law index of the mass accretion rate scaling with stellar mass.
    \item In the theoretical model of passive dust sublimation, $k$ represents the power-law index of the luminosity\text{--}mass dependence.
\end{enumerate}
}
\end{table*}

\begin{figure}
\centering
\includegraphics[width=\columnwidth]{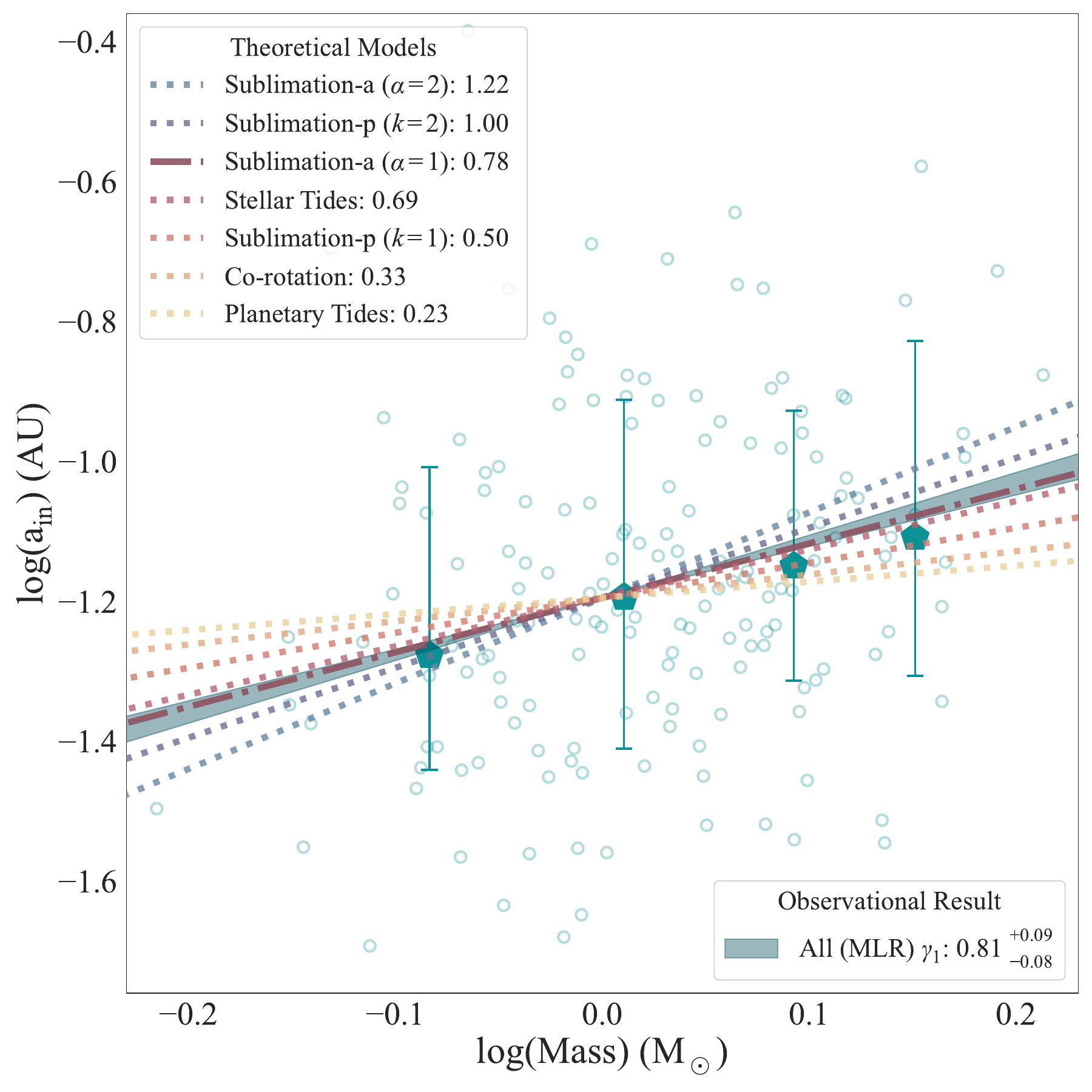}
\caption{Correlation $\gamma_1$ comparison between different theoretical models and the observational result. The green band represents the result of the dataset’s all multiple systems population obtained through the MLR model (same as the top panel in Fig. \ref{figml}, with points having the same meaning). The dotted and dotted-dashed lines in different colors represent the power-law indices under various theoretical models, arranged in descending order from top to bottom. For comparison of correlation $\gamma_1$, the intercepts of the models are standardized with the intercept of the observational result. 
\label{figallcompared}}
\end{figure}

\subsection{Implications for planet formation and evolution}
\label{sec:comparedmodels}

By conducting the above analysis on the dataset across four populations (all multiple systems, super-Earth systems, sub-Neptune systems, and mixed multiple systems), we generally confirm the results of previous studies \citep{2013ApJ...769...86P,2015ApJ...798..112M} and find that the relationship between the inner edge and stellar mass remains qualitatively consistent, with the inner edge position increasing with stellar mass. However, quantitatively, the degree of dependence between the inner edge and stellar mass appears to vary across different populations. Furthermore, by isolating the effect of stellar mass through correcting for metallicity, the influence of stellar mass on the inner edge is further strengthened, showing a correlation index \re{($\gamma_1$)} of $0.6-1.1$ with a power-law relationship (Figs. \ref{figdataset1correlation}, \ref{figml}, and Table \ref{tab:powerlawindex}). Additionally, observational selection bias does not impact our results. In the following, we discuss the implications of our results for planet formation and evolution.

There are a number of theoretical models in the literature exploring the relationship between the inner edge and stellar mass. For the pre-main-sequence dust sublimation radius of an actively accreting disk, as derived from Eq. (9) in \cite{2019A&A...632A...7L}, the relationship is given by $a_{\text{in}} \propto M_\star^{(3 + 4\alpha) / 9}$, where $\alpha$ represents the power-law index of the scaling between the mass accretion rate and stellar mass ($\dot{M}_{\text{acc}} \propto M_\star^{\alpha}$). Observations suggest $\alpha = 2$ \citep{2014A&A...561A...2A}, and we also examined the case where $\alpha = 1$. The corresponding power-law indices are $a_{\text{in}} \propto M_\star^{11/9}$ and $a_{\text{in}} \propto M_\star^{7/9}$, respectively. For the pre-main-sequence dust sublimation radius of a passive disk, $a_{\text{in}} \propto L_\star^{1/2}$ \citep{2001ApJ...560..957D,2017ApJ...845...44T} and $L_\star \propto M_\star^{k}$, where $k$ ranges from one to two during the pre-main-sequence stars. At earlier stages ($t < 1 \, \text{Myr}$), $k$ is approximately two, gradually declines toward one by $t \sim 10 \, \text{Myr}$. Thus, we considered two values for $k$: one and two. In that case, $a_{\text{in}} \propto M_\star^{0.5}$ for $k = 1$, and $a_{\text{in}} \propto M_\star^{1.0}$ for $k = 2$, based on the temperature profile of the minimum mass solar nebula. For planet destruction by stellar tides and by planetary tides maintained through secular interactions, the scaling follows Eqs. (15) and (16) in \cite{2015ApJ...798..112M}, which are derived from Eq. (1) in \cite{2009ApJ...698.1357J}. Assuming $R_\star \propto M_\star$, the respective power-law indices are $a_{\text{in}} \propto M_\star^{9/13}$ and $a_{\text{in}} \propto M_\star^{3/13}$. The pre-main-sequence co-rotation radius represents the location where the inner edge truncates at a frequency of planetary revolution synchronized with the stellar rotation frequency \re{\citep{2016A&A...591A..45P,2017ApJ...842...40L}}. In a state of Keplerian rotation, a relationship between stellar mass and the inner edge position is obtained as $a_{\text{in}} \propto M_\star^{1/3}$ \citep{2015ApJ...798..112M}.

Table \ref{tab:powerlawindex} summarizes these theoretical models, along with the observational results from previous work and this study. Figure \ref{figallcompared} illustrates the comparison between observational result and the theoretical models described above, using the same intercept (here, the intercept obtained from the MLR model for all multiple systems, \re{$\log(\gamma_{0,amf}) = -1.195$} in Eq. (\ref{lrlogall})). In Fig. \ref{figallcompared}, we observe that the Sublimation (active, ${\mathrm{\alpha}} = 1$) model ($\sim \re{0.78}$) provides the best fit, matching the observed correlation slope (\re{$0.81_{-0.08}^{+0.09}$}) within one sigma consistency at the same intercept. Additionally, the Stellar Tides model is also close to the observed result. However, the Sublimation (active, ${\mathrm{\alpha}} = 2$) model and the Sublimation (passive, $k = 2$) model show larger slopes, while the Sublimation (passive, $k = 1$) model, the Co-rotation Radius model, and the Planetary Tides model show smaller slopes.

In summary, dust sublimation is likely a key factor shaping the dependence of the inner edge of small planets on stellar mass. In addition, for the $\gamma_1$ values across all populations, which fall within the range of approximately $0.6-1.1$, we propose that both dust sublimation and stellar tides, having similar effects, could play a role in shaping this dependence. If these two mechanisms simultaneously influence the inner edge, their combined action might result in a stronger correlation. In future work, more theoretical models will be needed to explore the outcomes of these two mechanisms acting together. \re{Given that existing models can provide quantitative relationships between stellar mass and the inner edge during the pre-main-sequence phase, this study focuses on theoretical models related to planet formation. However, we emphasize that the evolution of planetary systems can also play a significant role. We believe that the observational results likely reflect the combined effects of both planetary system formation and evolution (see Sect. \ref{futurework}). Therefore, in the models discussed, tidal effects—such as stellar tides and planetary tides (i.e., evolutionary models)—should also be considered. However, it is important to note that this process primarily contributes to the formation of ultra-short-period planets, which have already been excluded from our study (see Sect. \ref{sec:sampleselection}).}

\subsection{Comparisons with previous studies}
\label{Differences}

\begin{figure}
\centering
\includegraphics[width=\columnwidth]{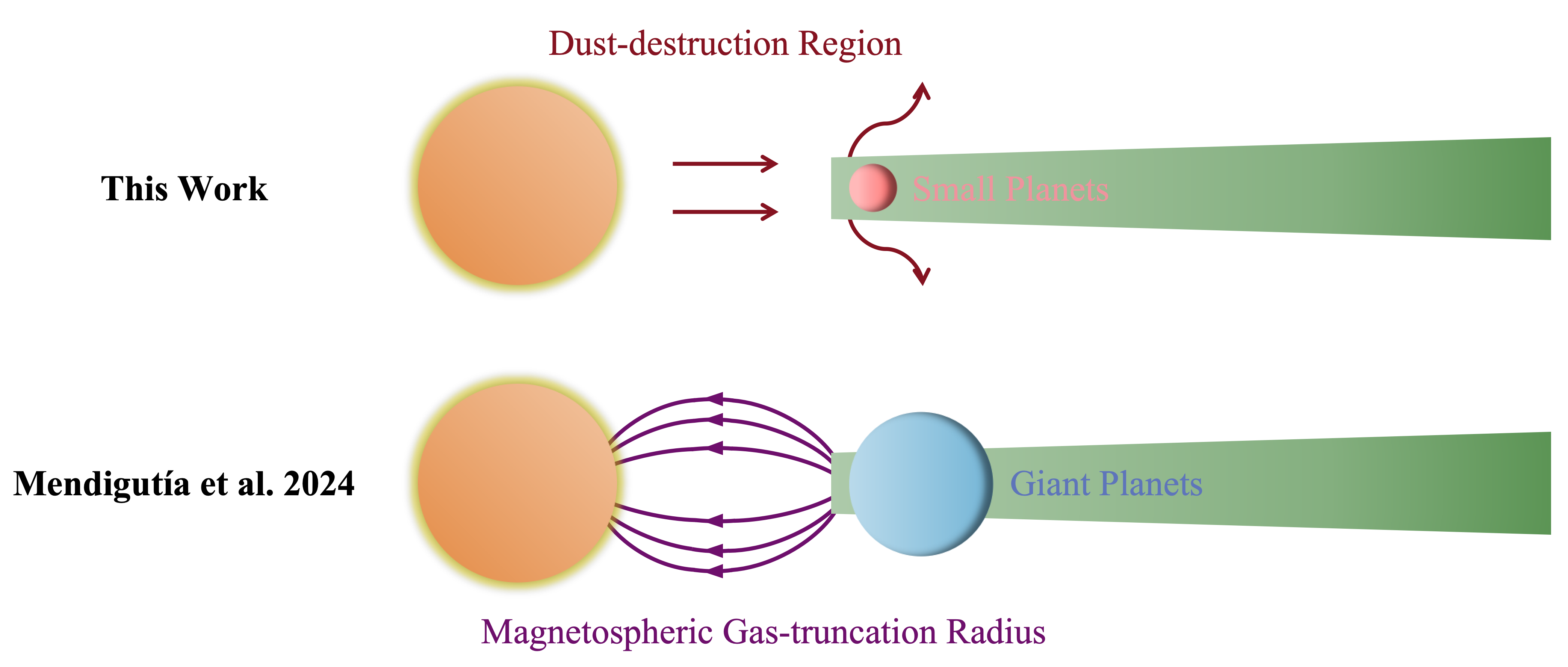}
\caption{Schematic comparison of the conclusions of this work and \cite{2024A&A...686L...1M}, revealing that the inner edges of different planetary systems correspond to different mechanisms. The results obtained in this work for small planets align more closely with the dust sublimation model, whereas their results obtained for giant planets are more consistent with migration up to the gas-truncation radius.
\label{figposter}}
\end{figure}

\cite{2024A&A...686L...1M} combined the Transiting Exoplanet Survey Satellite (TESS) and $Gaia$ DR3 data to study the location of short-period hot Jupiters and found that these planets around intermediate-mass stars tend to orbit closer to their host stars than the inner dust disk, aligning instead with the magnetospheric truncation radius. This suggests that the inner gas disk, rather than the dust disk, constrains the innermost orbits of hot Jupiters in such systems. Based on the above research and comparative analysis, we find that the inner dust disk constrains the innermost orbits of small planets, in contrast to the inner edges of giant planets, which are associated with the magnetospheres of protoplanetary disks \citep{2024A&A...686L...1M}. \re{It is important to note that their study employs a different research methodology from ours. Different objectives require different analytical approaches. Here, we primarily conduct a preliminary discussion and comparison of the conclusions drawn from the two studies.} We believe this comparative result is reasonable and have presented a schematic comparison in Fig. \ref{figposter}. For giant planets, their masses are primarily dominated by gas, and their migration occurs within the gas disk, making the inner edge dependent on the magnetospheric gas-truncation radius. In contrast, small planets are mainly composed of solid material, and their formation relies on the dust disk, resulting in the inner edge being determined by the dust-destruction region. Therefore, the comparison between this study and previous research indicates that the inner edges of different planetary populations (small planets and giant planets) are governed by distinct mechanisms. In summary, the inner edge for giant planets is likely determined by the migration halting mechanism, while for small planets, it is associated with the local solid distribution.

\re{While the alignment between the inner disk edge and the dust sublimation radius diminishes the necessity of orbital migration for small planets, migration cannot be entirely discounted provided that it ceases effectively near the sublimation boundary. For example, \cite{2024MNRAS.528.2883Z} proposed that the inner dead zone edge, rather than the magnetospheric truncation radius, determined a planet’s final position before the protoplanetary disc dissipated. As illustrated in Fig. 23 of their work, this effect appeared to be more pronounced for low-mass planets. While high-mass planets might still have migrated inward beyond the dust sublimation radius, low-mass planets were suggested to be unable to migrate within the inner magneto-rotational instability (MRI)-active region. Instead, they were likely to be trapped at the dead zone, which nearly coincided with the dust sublimation radius—where dust sublimated and MRI could operate. Thus, our observational results provide support for their theoretical simulation model, as the findings appear to be broadly consistent with our observations.}

\cite{2013ApJ...769...86P} used short-period confirmed Jovian exoplanets and $Kepler$ planetary candidates from the early data release to study the relationship between the semimajor axis of all planets and stellar mass (planet frequency as a function of stellar mass). They found that the tidal halting mechanism provided the best fit for confirmed Jovian exoplanets. However, for $Kepler$ candidates, they did not confirm a specific mechanism but noted that the power-law index $\gamma_1 = 0.38-0.9$ is larger than the predicted range for the tidal halting mechanism ($0.23-0.33$). In contrast, by using only the innermost planet in multis as the inner edge to study its relationship with stellar mass, we find a power-law index of $\gamma_1 = 0.6-1.1$. This result is comparable to \cite{2013ApJ...769...86P}’s finding for $Kepler$ candidates, although the samples are not entirely the same.

\cite{2015ApJ...798..112M} studied sub-Neptunes and super-Earths ($1 \, R_\oplus$ to $4 \, R_\oplus$) to investigate the occurrence of small planets around stars of different masses. They found an occurrence drop at an orbital period of approximately 10 days, regardless of stellar mass. Using this occurrence drop to trace the inner edge of planetary systems, they derived a relationship of $a_{\text{in}} \propto M_\star^{1/3}$, suggesting that the co-rotation radius best matches the stellar-mass-dependent location of the inner edge. However, by using the innermost planets in multis as the inner edge to study the relationship with stellar mass, we find a stronger correlation with stellar mass (around $a_{\text{in}} \propto M_\star^{0.6-1.1}$), which aligns well with the prediction of dust sublimation model. We find that the discrepancy stems from using planet occurrence rates to trace the inner edge, which underestimates the correlation index ($\gamma_1$) and diminishes the influence of stellar mass on the location of inner-edge planets. A more detailed discussion and analysis is provided in appendix \ref{Mulders2015}. Previous studies analyze the position of the inner edge using planet occurrence rates \citep[such as Eqs. (7), (8), and (9) in][]{2015ApJ...798..112M}, which consider all planets in the system. In this method, outer planets will have more weight (due to the correction of the transit geometric effect) to determine the inner edge than the innermost one, although the innermost one represents the true inner edge. Consequently, in multis, lower-mass stars tend to host more planets \citep{2020AJ....159..164Y}, shifting the inner edge derived from occurrence rates outward. This shift results in a weaker correlation between the inner edge and stellar mass compared to the intrinsic correlation. As shown in appendix \ref{Mulders2015}, if the intrinsic correlation between the inner edge and stellar mass has a power-law index of 1.0, it would be underestimated as 0.3 when using the occurrence rate method. This effectively explains the discrepancy between the results of \cite{2015ApJ...798..112M} and this work.

Additionally, \cite{2012ApJS..201...15H}, \cite{2015ApJ...798..112M}, \cite{2020AJ....159..164Y}, and \cite{2025AJ....169...45G} have all found that close-in planets are rarer around higher-mass stars, a result qualitatively consistent with that as stellar mass increases, the position of the inner edge shifts outward. Compared to FGKM-type stars, A-type stars have significantly higher temperatures (with more pronounced differences in temperature than stellar mass). As a result, the dust sublimation radius shifts outward, pushing the inner edge farther away and leaving little room for small, close-in planets to form. Using the relationship between the inner edge and stellar mass derived in this work (All-MLR-$\gamma_1$: \re{$0.81_{-0.08}^{+0.09}$}), we extrapolate to A-type stars ($\sim 2 \, M_{\odot}$) and find that the inner edge would reach $0.21-0.27$ AU. This implies that it is challenging for planets to exist within 0.2 AU around A-type stars, which aligns well with the results of \cite{2025AJ....169...45G}, demonstrating that close-in planets orbiting A-type stars are rare.

\subsection{Inspirations for future work}
\label{futurework}

Although the observational results generally align with some of the theoretical models in Fig. \ref{figallcompared}, we can see that there is a significant scatter in the observational data, reflected in large uncertainties in the intercepts. The intercept essentially acts as a normalization factor \re{(e.g., $\log(\gamma_{0,am})$ in Eq. (\ref{logaM}) and $\log(\gamma_{0,amf})$ in Eq. (\ref{lrlogall})).} This may be related to two factors: the surface density of the protoplanetary disk and the size of the grains or dust within the disk \citep[see Fig. 8 in][]{2009A&A...506.1199K}. Thus, the large scatter of observational data around theoretical models (Fig. \ref{figallcompared}) may reflect the diversity in the initial conditions of the protoplanetary disk. Furthermore, the relationship between the inner edge and stellar mass may be influenced by more complex factors. For instance, some studies \citep{2017A&A...601A..15L,2017A&A...606A..66L} suggest that after planet formation, the disk dissipation process further affects the position of the inner edge, as the orbits of planets expand with the inner magnetospheric cavity’s growth. In addition, other dynamical processes at later stages, such as giant impacts \citep{2017MNRAS.470.1750I,2023AREPS..51..671G}, planetary scattering \citep{2008ApJ...686..621F,2008ApJ...686..580C}, and secular chaos \citep{1997A&A...317L..75L,2011ApJ...739...31L}, may further shape the architecture of planetary systems, changing the position of the innermost planets. \re{Future work could distinguish between resonant and nonresonant systems to further explore the relationship between the inner edge and stellar parameters. In this context, we emphasize the potential role of evolutionary processes. However, studying such effects requires precise stellar age measurements, which remain challenging due to large uncertainties in current data. Despite these limitations, understanding the evolutionary history of planetary systems remains an important goal.} Therefore, future theoretical model simulations should take into account more factors, as the strength of the scatter is influenced by multiple elements. Exploring the combinations of these factors and their combined effects would be a direction for further research.

In this work, focusing on small planets, we find that different populations exhibit varying correlations between the inner edge and stellar mass (Table \ref{tab:powerlawindex}). Among them, mixed multiple systems show the strongest correlation, while super-Earths and sub-Neptunes exhibit much weaker correlations. This may be because small planets are inherently diverse and complex, with different populations undergoing distinct formation and evolution processes that may correspond to different theoretical mechanisms. However, the sample size in this work is still relatively small. Future studies with more data are needed to further validate and explore these findings. \re{Additionally, to assess the impact of radius uncertainties on planet classification, we conduct simulations in which planets are repeatedly reclassified and refitted. The resulting slope distributions for super-Earths and sub-Neptunes are $0.57_{-0.10}^{+0.10}$ and $0.45_{-0.17}^{+0.17}$, respectively. These results show increased uncertainties compared to the observed values of $0.54_{-0.05}^{+0.05}$ and $0.38_{-0.09}^{+0.09}$, which remain consistent within $1\sigma$. Thus, incorporating radius uncertainties enlarges the error bars and reduces the statistical significance of the difference between the two populations. Given the small sample sizes, this difference remains suggestive rather than conclusive.}

Apart from stellar mass, which is the primary focus of this work, the inner edge is likely determined by multiple properties. The correlation between the inner edge and stellar mass may reflect the role of dust sublimation in shaping the inner edge. However, the theoretical understanding of how other properties, such as stellar metallicity, influence the inner edge remains unclear. This could also contribute to the significant scatter observed. In addition, stellar age is another important property. Planetary systems evolve over time, and stellar age is also correlated with other factors. Therefore, analyzing the influence of stellar age is crucial for understanding the long-term evolution of planetary systems. In the future, further theoretical and observational work is needed to fully understand the influence of properties such as stellar metallicity and stellar age on the inner edge. \re{In addition, larger planets also represent an interesting research direction. However, since this study already covers an extensive scope and categorizes the sample into super-Earths and sub-Neptunes, we do not further explore the impact of larger planets in this work. Nonetheless, we suggest that future research incorporate planetary mass or radius into the relationship between the inner edge and stellar mass for a more comprehensive analysis.}

The edges of planetary systems are essential for understanding the mechanisms of planet formation. While this work focuses on the inner edge, it is important to note that the outer edge is equally significant. \cite{2022AJ....164...72M} and \cite{2023ApJ...954..137S}, in their studies of the outer edges of $Kepler$ multis, conducted detailed analyses and demonstrated that the existence of the outer edge is not a result of observational limitations, but rather due to specific physical mechanisms. Nonetheless, the theoretical truncation mechanism responsible for the outer edge remains unknown. Therefore, we are curious to explore whether there is any connection between the inner edge and the outer edge. 

\section{Conclusion and summary}
\label{sec:sumandcon}

The position of the innermost planet in a planetary system reflects the relationship between the entire system and its host star, offering potentially significant insights into planetary formation and evolution processes. However, previous studies have revealed persistent discrepancies between observational results and theoretical models, leaving the primary mechanism governing the position of the inner edge in planetary systems uncertain. In this work, we used the $Kepler$ DR25 catalog combined with LAMOST and $Gaia$ data and focused specifically on small planets in multis (Table \ref{tab:dataselection} and Fig. \ref{figdata2HRandPR}). The data were categorized into four planetary populations (Fig. \ref{figKBPin}): all multiple systems, super-Earth systems, sub-Neptune systems, and mixed multiple systems. Our aim was to study the correlation between stellar mass and the position of the inner edge. 

Our findings revealed that as stellar mass increases, the position of the inner edge also increases (Fig. \ref{figdataset1correlation}). Simultaneously, we observed that the inner edges of hot small-planet systems decrease with stellar metallicity (Fig. \ref{figdata2fehcorrelationmul}). Since stellar mass is also correlated with metallicity, the intrinsic correlation between the inner edge and stellar mass should therefore be even stronger. After correcting for metallicity dependence using an MLR model, we observed a steeper correlation between stellar mass and the inner edge (Fig. \ref{figml}). Analyzing the dataset across four planetary populations revealed a correlation power-law index of $\gamma_1 = 0.6-1.1$. In addition, we found that observational biases do not affect our results (Fig. \ref{figdata2obsselectmul}).

In analyzing the observational data, we identified that using occurrence rates, as employed in previous studies, introduces a bias that underestimates the dependence of inner-edge planets on stellar mass (appendix \ref{Mulders2015}). Correcting for this influence reveals a stronger relationship between the inner edge and stellar mass. This effect explains the discrepancy between this work ($\gamma_1 = 0.6-1.1$) and a previous study \citep[$\gamma_1 = 0.3$,][]{2015ApJ...798..112M}.

By comparing our results with existing theoretical models (Table \ref{tab:powerlawindex} and Fig. \ref{figallcompared}), we found that the observed correlation between the inner edge and stellar mass aligns with the pre-main-sequence dust sublimation radius of the protoplanetary disk. This supports the hypothesis that dust sublimation is an important mechanism that affects the formation and orbital evolution of small planets \citep{2013arXiv1306.0566B,2016ApJ...827..144F}, contrasting with the inner edges of hot Jupiters, which are influenced by the magnetospheres of protoplanetary disks \citep{2024A&A...686L...1M}. This underscores the possibility that the inner edge of different planetary populations may indeed be governed by distinct mechanisms (Fig. \ref{figposter}).

It is important to note that although the observational trend best fits the dust sublimation model, we caution that there remains significant scatter in the observational data around the best-fit model. This is likely due to the complex and multifaceted factors influencing the correlation between the inner edge and stellar mass, such as disk density and grain size, the inner magnetospheric cavity during the disk dissipation process, giant impacts, planetary scattering, and secular chaos. Further research is needed to better understand and address these issues in the future.

\begin{acknowledgements}
      We thank Qi Hao, Jiayi Yang, and Shuting Han for their valuable discussions on methods, analyses, and figures. We also extend our gratitude to the referee for their insightful comments. \ree{This work is supported by the National Key R\&D Program of China (No. 2024YFA1611803) and the National Natural Science Foundation of China (NSFC; Grant Nos. 12273011, 12150009, 12222303, and 12173035). J.-W. X. also acknowledges the support from the National Youth Talent Support Program. B. L. is also supported by the start-up grant of the Bairen program from Zhejiang University. Funding for LAMOST (\url{www.lamost.org}) has been provided by the Chinese NDRC. LAMOST is operated and managed by the National Astronomical Observatories, CAS.}  
\end{acknowledgements}

\bibliographystyle{aa}
\bibliography{aa53671-25}

\begin{appendix}
\onecolumn
\section{Occurrence rates underestimate the inner edge \text{--} stellar mass correlation}
\label{Mulders2015}

In this appendix, we demonstrate that using planet occurrence rates weakens the observed dependence of inner-edge planets on stellar mass. Qualitatively, the calculation of occurrence rates does not exclusively consider the innermost planet in a planetary system; instead, all planets contribute to the result. In single-planetary systems, the position of the planet directly represents the inner edge. Therefore, occurrence rates can accurately reflect the inner edge in such systems. However, in systems with additional outer planets, the inner edge derived from occurrence rates is consistently shifted outward. Compared to the innermost planet, outer planets generally have lower transit probabilities and thus contribute more weight to the calculation of occurrence after correcting the transit observational bias. As a result, systems with more planets exhibit a more pronounced outward shift in the calculated inner edge. Since the number of planets in a system decreases with increasing stellar mass \citep{2020AJ....159..164Y}, systems around lower-mass stars, which typically host more planets, experience a greater outward shift in the inner edge. In contrast, higher-mass stars, with fewer planets, exhibit a less significant shift. This variation in outward shift offsets part of the intrinsic relationship between the inner edge and stellar mass, ultimately weakening the observed correlation. 

\begin{figure*}[htbp]
\centering
\includegraphics[width=0.79\textwidth]{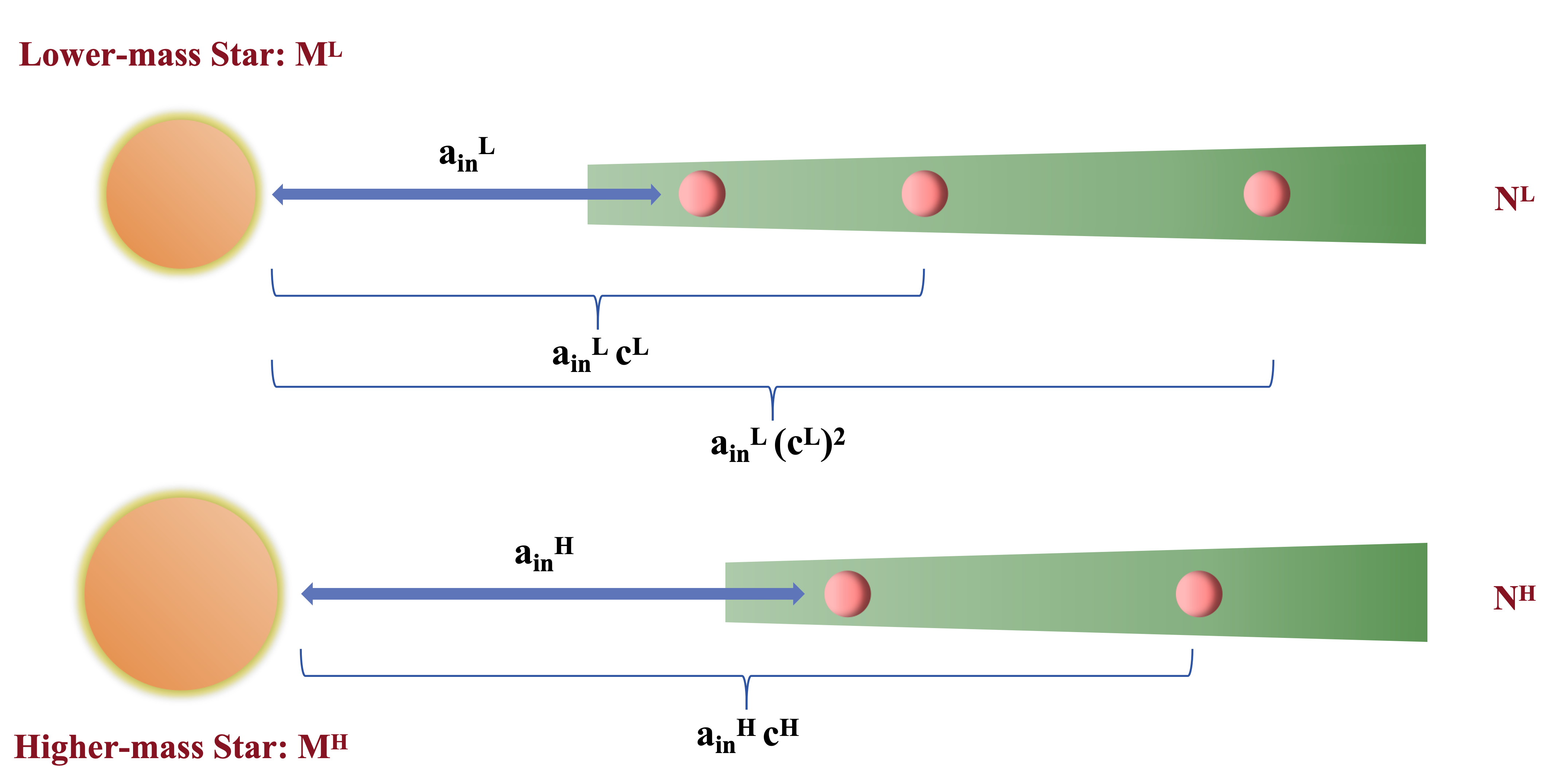}
\caption{Schematic diagram illustrating the impact of planetary occurrence rates on the conclusion. In the diagram, $M^\text{L}$ and $M^\text{H}$ represent lower-mass and higher-mass host stars, respectively, while $N^\text{L}$ and $N^\text{H}$ denote the number of planets in each system ($N^\text{L} > N^\text{H}$). $a_\text{in}^\text{L}$ and $a_\text{in}^\text{H}$ represent the semimajor axes of the innermost planets in the two systems. $c^\text{L}$ and $c^\text{H}$ represent the conversion factor between the period ratio and the semimajor axis ratio of the planetary distribution in the system.
\label{figocc}}
\end{figure*}

We further conduct a quantitative analysis to specifically calculate the extent to which the relationship between the inner edge and stellar mass has been weakened. According to equations from \cite{2015ApJ...798..112M}, the occurrence rate for a planet of a given radius and orbital period is 
\begin{equation}
f_{\text{occ}}(\{T_{\text{eff}}\}, R_{\text{p}}, P) = \frac{1}{f_{\text{geo}} N_{\star}(\{T_{\text{eff}}\}, R_{\text{p}}, P)},
\label{focc}
\end{equation}
where 
\begin{equation}
f_{\text{geo}} = \frac{R_{\text{p}} + R_{\star}}{a(1 - e^2)} \, , \;
N_{\star} (\{T_{\text{eff}}\}, R_{\text{p}}, P) = \sum_{i=0}^{N_{\star}(\{T_{\text{eff}}\})} \left( f_{\text{eff}, \text{i}} \cdot f_{\text{n}, \text{i}} \right) \, , \;
\end{equation}
and $N_{\star}$ is rounded to the nearest integer. Considering that the $Kepler$ system we are studying consists of nearly circular orbits ($e = 0$) and $R_{\text{p}} \ll R_{\star}$, for a given system, $R_{\star}$ and $N_{\star}$ remain consistent. Therefore, we can derive that $f_\text{occ} \propto a_\text{occ}$. Following \cite{2015ApJ...798..112M}, the inner edge calculated from occurrence rates ($a_\text{occ}$) is defined at the point where $f_\text{occ}$ drops by half (or the mean point). Thus, we can derive 
\begin{equation}
f_{\text{occ}} \propto a_{\text{occ}} = \frac{\sum_{\text{n}=1}^{N} \left(a_{\text{n}} \right)}{N}, 
\label{occain}
\end{equation}
where $a_\text{occ}$ represents the inner edge of a planetary system calculated through $f_\text{occ}$, $N$ represents the number of planets in the system, and $a_{\text{n}}$ denotes the semimajor axis of each planet in the system, ordered from innermost to outermost. We assume that the semimajor axis ratio between two adjacent planets in a system is constant and related to their period ratio, which can be calculated through $a = \sqrt[3]{GM_{\star} P^2 / 4\pi^2}$. Thus, we can further derive 
\begin{equation}
f_\text{occ} \propto a_\text{occ} = \frac{a_\text{in} \cdot \sum_{\text{n}=0}^{N-1} c^{\text{n}}}{N},
\label{occainc}
\end{equation}
where $a_\text{in}$ represents the real position of the innermost planet, and $c = a_{\text{n}+1} / a_{\text{n}} = \left(P_{\text{n}+1} / P_{\text{n}} \right)^{2/3}$ represents the conversion process between the period ratio and the semimajor axis ratio for planets in the system.

\begin{figure*}[htbp]
\centering
\includegraphics[width=\textwidth]{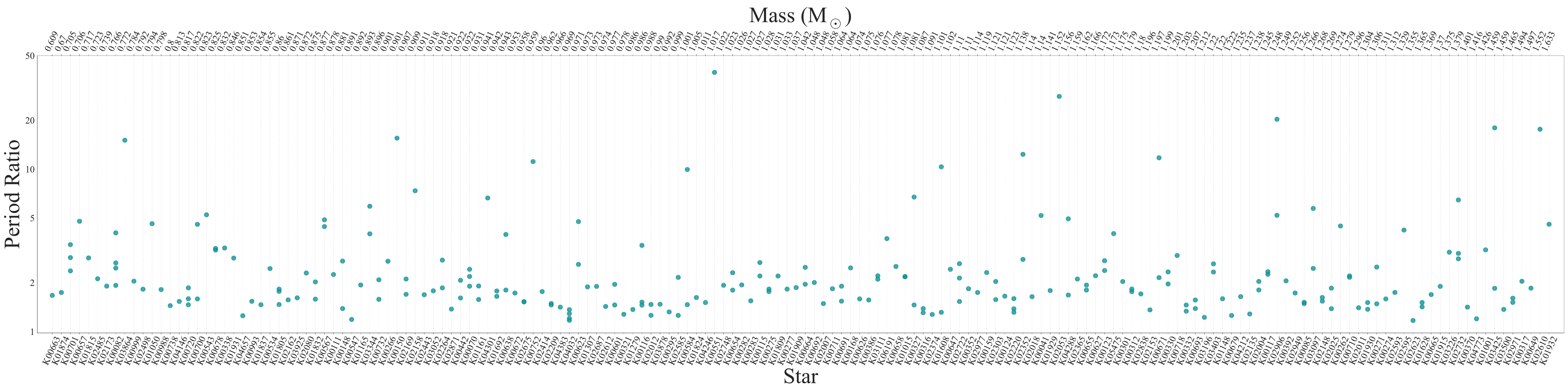}
\caption{Period ratios of all the systems studied in this work. From left to right, the host star mass of the planetary systems gradually increases.
\label{figc}}
\end{figure*}

To facilitate quantitative calculations and to visually observe the impact of planet occurrence rates, we hypothesize two planetary systems, as shown in Fig. \ref{figocc}. $M^\text{L}$ and $M^\text{H}$ represent lower-mass and higher-mass host stars, respectively, while $N^\text{L}$ and $N^\text{H}$ denote the number of planets in each system ($N^\text{L} > N^\text{H}$). $a_\text{in}^\text{L}$ and $a_\text{in}^\text{H}$ represent the semimajor axes of the innermost planets in the two systems. $c^\text{L}$ and $c^\text{H}$ represent the semimajor axis ratios of adjacent planets in two respective systems. For these two systems, we can derive the following formulas:
\begin{equation}
f_\text{occ}^\text{L} \propto a_\text{occ}^\text{L} = \frac{a_\text{in}^\text{L} \cdot \sum_{\text{n}=0}^{N^{\text{L}}-1} (c^\text{L})^{\text{n}}}{N^\text{L}} \, , \;
f_\text{occ}^\text{H} \propto a_\text{occ}^\text{H} = \frac{a_\text{in}^\text{H} \cdot \sum_{\text{n}=0}^{N^{\text{H}}-1} (c^\text{H})^{\text{n}}}{N^\text{H}}.
\end{equation}
Therefore, 
\begin{equation}
\frac{a_{\text{occ}}^{\text{H}}}{a_{\text{occ}}^{\text{L}}} = \frac{a_{\text{in}}^{\text{H}}}{a_{\text{in}}^{\text{L}}} \cdot \frac{N^{\text{L}} \cdot \sum_{\text{n}=0}^{N^{\text{H}}-1} (c^{\text{H}})^{\text{n}}}{N^{\text{H}} \cdot \sum_{\text{n}=0}^{N^{\text{L}}-1} (c^{\text{L}})^{\text{n}}}.
\label{occin}
\end{equation}
In this work, we express the correlation between the inner edge and stellar mass using $a_\text{in} \propto M_{\star}^{\gamma_1}$ and $a_\text{occ} \propto M_{\star}^{\gamma_\text{occ}}$; then we have 
\begin{equation}
\frac{a_{\text{occ}}^{\text{H}}}{a_{\text{occ}}^{\text{L}}} = (\frac{M^{\text{H}}}{M^{\text{L}}})^{\gamma_\text{occ}} \, , \;
\frac{a_{\text{in}}^{\text{H}}}{a_{\text{in}}^{\text{L}}} = (\frac{M^{\text{H}}}{M^{\text{L}}})^{\gamma_1}.
\label{occandin}
\end{equation}
Thus, through Eq. (\ref{occandin}), Eq. (\ref{occin}) can be transformed into
\begin{equation}
\left(\frac{M^{\text{H}}}{M^{\text{L}}}\right)^{\gamma_{\text{occ}} - \gamma_1} = \frac{N^{\text{L}} \cdot \sum_{\text{n}=0}^{N^{\text{H}}-1} (c^{\text{H}})^{\text{n}}}{N^{\text{H}} \cdot \sum_{\text{n}=0}^{N^{\text{L}}-1} (c^{\text{L}})^{\text{n}}}.
\label{moccin}
\end{equation}
By converting into the logarithmic form, we can derive the final analytical expression:
\begin{equation}
\Delta \gamma = \gamma_{\text{occ}} - \gamma_1 = \log_{\left(\frac{M^{\text{H}}}{M^{\text{L}}}\right)} \left(\frac{N^{\text{L}} \cdot \sum_{\text{n}=0}^{N^{\text{H}}-1} (c^{\text{H}})^{\text{n}}}{N^{\text{H}} \cdot \sum_{\text{n}=0}^{N^{\text{L}}-1} (c^{\text{L}})^{\text{n}}}\right).
\label{mgamma}
\end{equation}
Here, we analyze the period ratios of all the systems studied in this work, from which we can derive the corresponding $c$ for each system. Based on the period ratio and stellar mass data shown in Fig. \ref{figc}, we calculate the average period ratios for systems with stellar masses less than $1 \, M_{\odot}$ and greater than or equal to $1 \, M_{\odot}$, which are 2.894 and 2.827, respectively. So we can assume that the $c$ value is consistent between higher-mass and lower-mass star systems (i.e., ${c^{\text{L}}} \sim {c^{\text{H}}} = 2.8^{2/3} \sim 2$).

Using Eq. (\ref{mgamma}), we can analyze that $\Delta \gamma$ represents the extent to which the real inner edge is underestimated when calculated through the occurrence rate. If the number of planets in the systems of low-mass and high-mass stars is the same ($N^\text{L} = N^\text{H}$), it is evident that $\Delta \gamma = 0$. Thus, the correlation between the inner edge and stellar mass is not underestimated. However, if we assume the mass of the higher-mass star is $M^\text{H} = 1.0 \, M_{\odot}$, and the mass of the lower-mass star is $M^\text{L} = 0.5 \, M_{\odot}$. According to the results of \cite{2020AJ....159..164Y}, systems with $1.0 \, M_{\odot}$ host stars have an average of two planets ($N^\text{H} = 2$), while systems with $0.5 \, M_{\odot}$ host stars have an average of three planets ($N^\text{L} = 3$). Thus, 
\begin{equation}
\Delta \gamma = \gamma_{\text{occ}} - \gamma_1 = \log_2 \left(\frac{3 \cdot \left(1 + c^{\text{H}}\right)}{2 \cdot \left(1 + c^{\text{L}} + (c^{\text{L}})^2\right)}\right) \approx \re{-0.64}.
\end{equation}
It is evident that the relationship between the true inner edge and stellar mass has been significantly underestimated. Based on our conclusion that the power-law index of the correlation between the inner edge and stellar mass, we approximately assume $\gamma_1 = 1.0$. Thus, we obtain $\gamma_{\text{occ}} = \re{0.36}$, which aligns well with the conclusion in \cite{2015ApJ...798..112M}, approximately 0.3.

Therefore, we find that, compared to the results obtained from the true inner edge, the relationship between the inner edge and stellar mass derived through the occurrence rate calculation method is weakened. This explains why the conclusion of \cite{2015ApJ...798..112M} shows a weaker correlation, whereas our findings reveal a stronger correlation ($\gamma_1 = 0.6-1.1$).
\end{appendix}

\end{document}